%% file: submission_arxiv.tex
\documentclass[twocolumn,showpacs,superscriptaddress,preprintnumbers,amsmath,amssymb]{revtex4-2}
\usepackage{graphicx}
\usepackage{dcolumn}
\usepackage{appendix}
\usepackage{bm}

\usepackage[usenames,dvipsnames]{xcolor}
\usepackage{subfigure}
\usepackage{bbm}
\usepackage{enumerate}
\usepackage[
  bookmarks=true,
  colorlinks,
  linkcolor=blue,
  urlcolor=blue,
  citecolor=blue,
  plainpages=false,
  pdfpagelabels,
  final,
  breaklinks=true
]{hyperref}

\usepackage{titlesec}
\usepackage{array}
\usepackage{dsfont}
\usepackage{physics}

\newcommand{\opE}{ {\vb{E}}_Q }
\newcommand{\pol}{ {\boldsymbol{\epsilon}} }
\newcommand{\Gone}{ G^{(1)} }
\newcommand{\Gtwo}{ G^{(2)} }

\begin{document}

\title{Theory of quantum optics and optical coherence in high harmonic generation}

\author{Philipp Stammer}
\email{philipp.stammer@icfo.eu}
\affiliation{ICFO-Institut de Ciencies Fotoniques, The Barcelona Institute of Science and Technology, Castelldefels (Barcelona) 08860, Spain.}
\affiliation{Atominstitut, Technische Universit\"{a}t Wien, 1020 Vienna, Austria}

\author{Javier Rivera-Dean}
\affiliation{ICFO-Institut de Ciencies Fotoniques, The Barcelona Institute of Science and Technology, Castelldefels (Barcelona) 08860, Spain.}

\author{Maciej Lewenstein}
\affiliation{ICFO-Institut de Ciencies Fotoniques, The Barcelona Institute of Science and Technology, Castelldefels (Barcelona) 08860, Spain.}
\affiliation{ICREA, Pg. Llu\'{\i}s Companys 23, 08010 Barcelona, Spain}

\date{\today}

\begin{abstract}

Optical coherence encodes information about the correlations of the electromagnetic field. In combination with quantum optical approaches, it allows for the study of the correlations between photons. 
Since the pioneering papers of Glauber, studies of optical coherence have  facilitated many fundamental insights into non-classical signatures of light emission processes, with wide applicability in modern quantum technologies.
However, when it comes to the photon up-conversion process of high-order harmonic generation the description has focused on semi-classical methods for decades. 
In this work, we overcome this limitation and establish a quantum optical theory of field correlations for the process of high harmonic generation (HHG). In effect, we introduce the notion of optical coherence at the intersection of quantum optics and strong laser-driven processes, and obtain the harmonic field correlation functions. 
In particular, we focus on the first and second order field correlation, which allow to understand the origin of the classical properties of the HHG spectrum, and its departure into the quantum regime. Further, we develop the theory for two-time intensity correlation functions of the harmonic field, and demonstrate the onset of anti-bunching signatures in HHG. 
We study the correlation functions in the regime of a single, few and many emitters in atomic HHG, showing the transition from quantum to classical signatures in the correlations. 
Since the theory is generic, it can be extended to multi-time correlation functions of any order, and allows to consider the interaction of light with arbitrary material systems.

\end{abstract}

\maketitle

\section{\label{sec:intro}Introduction}


Quantum optical coherence theory was formulated in the 1960s in a series of papers by Roy J. Glauber~\cite{glauber1963quantum, glauber1963states, glauber1963photon, glauber2006nobel}, who combined the wave-particle dualism of quantum electromagnetic (EM) fields, i.e. Maxwell's wave description and the explanation via photons. These papers gave birth to contemporary quantum optics, studying field properties in terms of correlation functions. Quantum optics (QO) studies EM field properties constituted out of individual photons, while field correlation functions are used to investigate the coherence properties of the radiation field~\cite{glauber1963quantum, mandel1995optical}. Of special interest are the first and second-order correlation functions, since they are associated with the spectrum and photon correlations of the field, respectively~\cite{carmichael2013statistical}.

A well studied paradigmatic example is the process of resonance fluorescence~\cite{kimble1976theory}, in which a two-level system (TLS) is driven by a classical field and scatters radiation into the vacuum. 
In view of a classical description, the driven emitter is associated with an induced dipole oscillator emitting coherent radiation, and the result agrees with the semi-classical picture for small driving intensities~\cite{heitler1984quantum}. However, the TLS is not an oscillator, and the quantum mechanical dipole moment is probabilistic such that averages over the dipole operator can acquire a stochastic component~\cite{carmichael2013statistical}. 
In other words, the classical and semi-classical pictures fail when looking carefully, i.e. when using the proper observables such as the statistics of the scattered photons. In particular, the semi-classical theory fails at relatively high driving laser intensities. Interestingly, at very high intensities the semi-classical theory becomes accurate again, as we will see below for the case of high-order harmonic generation~\cite{lewenstein1994theory}. 

\begin{figure*}
    \centering
	\includegraphics[width= \textwidth]{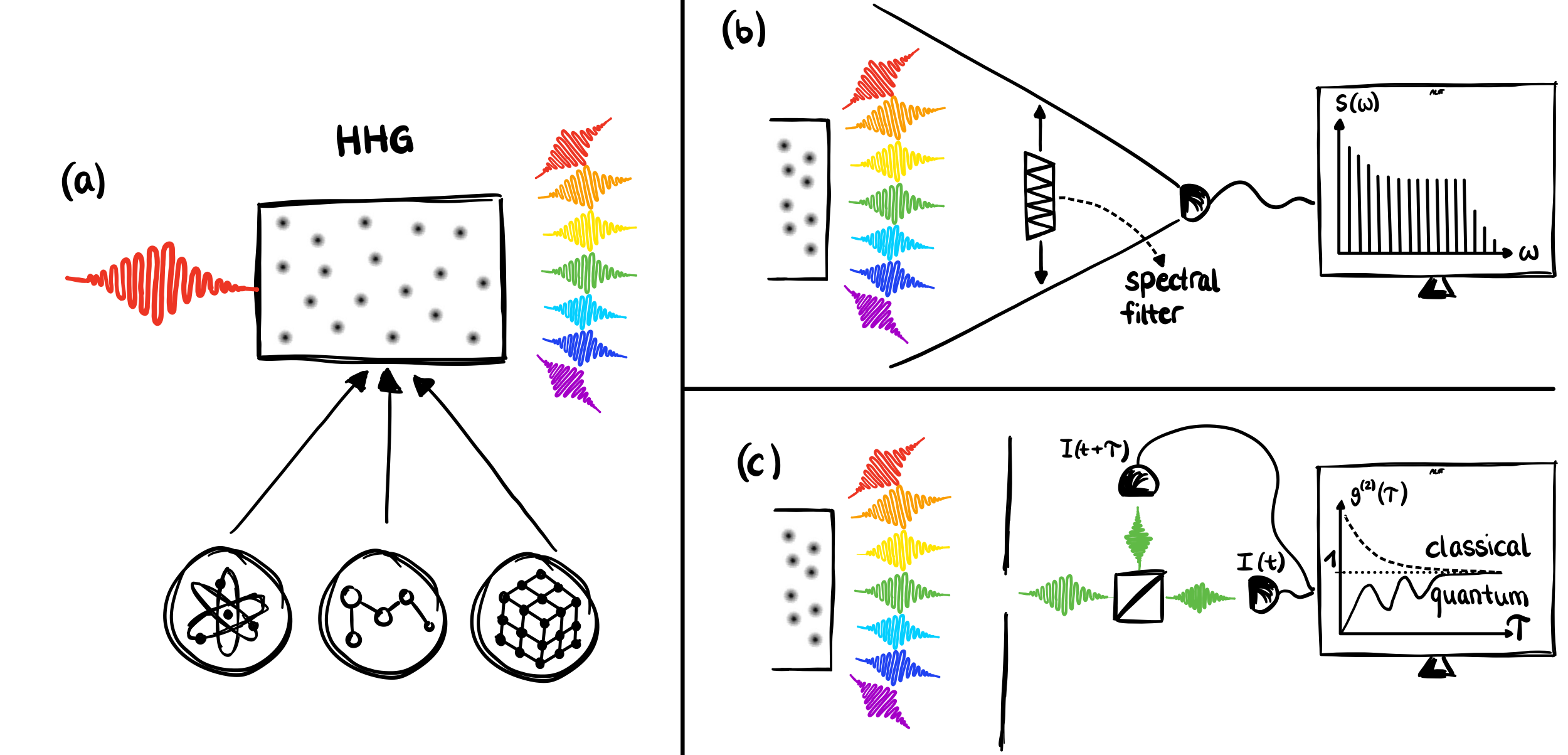}
	\caption{Here, we schematically show the experimental configurations studied in this work using the process of high harmonic generation (HHG). In (a) an illustration of the HHG process, in which an intense driving field induces highly non-linear electron dynamics in an atomic, molecular or solid state target leading to the light scattering event of HHG. The generated harmonics span over a wide frequency range up to the extreme-ultraviolet regime. In (b) and (c) we show the configuration for measuring the first and second order correlation function of the harmonics, respectively. In (b) the spectrum is recorded leading to the well known harmonic peaks, while in (c) the intensity correlation of a single harmonic shows anti-correlated photon emission.  }
      \label{fig:intro}
\end{figure*}

Coming back to the driven TLS, it was observed that for increasing laser intensity the spectrum splits into three peaks, known as the Mollow triplet~\cite{mollow1969power}, which originates from an additional incoherent contribution to the spectrum. This incoherent contribution arises from the quantum fluctuations of the dipole moment around the steady-state average, which are intrinsic to the probabilistic nature of the underlying quantum dynamics (note that the Mollow triplet can also be explained in the dressed-state picture~\cite{cohen1977dressed} due to the Autler-Townes effect~\cite{autler1955stark}). 
However, in addition to the resonance fluorescence spectrum, one can look at higher order correlation functions such as the second order intensity correlation~\cite{scully1997quantum}, as introduced in the seminal work by Hanbury Brown and Twiss (HBT) who measured photon bunching in thermal radiation in an intensity interferometer~\cite{brown1956correlation}. Measuring the intensity correlation at two detectors with a variable time-delay $\tau$ between the detector counts allows for inference about the temporal statistics of the emitted photons (see Fig.~\ref{fig:intro} for a schematic experimental configuration). In particular, the normalized second order intensity correlation function $g^{(2)}(\tau)$ is used to witness non-classical signatures in the photon correlations. For instance, the effect of photon anti-bunching is the case in which $g^{(2)}(0) < 1$, implying anti-correlations between photon counts at the two detectors, and the case of $g^{(2)}(0) = 0$ is evidence for the presence of only a single photon. In fact, the process of resonance fluorescence was used to show anti-bunching signatures~\cite{kimble1976theory}, representing the first evidence for the existence of optical photons~\cite{kimble1977photon}. 
An optical field is said to be non-classical if it shows anti-bunching signatures $g^{(2)}(0)< 1$ in the measured intensity correlations. The case of $g^{(2)}(0)=1$ is associated to classical coherent radiation, where the photons are not correlated and arrive at random~\cite{carmichael2013statistical}, while incoherent (classical) thermal light in contrast shows bunching properties in the intensity correlations~\cite{scully1997quantum}, given by $g^{(2)}(0) = 2$.

With the advent of high intensity laser sources of femtosecond pulse duration, and the associated non-linear dynamics of the driven emitter, the semi-classical picture experienced a resurgence~\cite{brabec2000intense}. In this regime of high intensity laser driven systems the classical and semi-classical picture of the light emission process resume its predictive power. With the observation of high-order harmonics of the driving laser frequency~\cite{ferray1988multiple, l1993high}, it was shown that classical methods can provide a good description of the observed harmonic spectrum~\cite{corkum1993plasma}. Furthermore, the subsequent semi-classical approach based on the dipole moment of the laser driven electron has been the theoretical building block to describe the process of high harmonic generation (HHG)~\cite{lewenstein1994theory}. This is achieved by considering the dipole moment expectation value of the coherently driven electron, while the non-linearity of the process is encoded in the highly anharmonic oscillation of the dipole. In contrast, early attempts to provide a full quantum optical description of the process of HHG has remained without much attention~\cite{guo1988quantum, xu1993non, compagno1994qed, becker1997unified, gauthey1995role}. On the one hand, this could be due to the non-perturbative nature of the HHG process~\cite{eberly1989high, eberly1989high2, eberly1989nonlinear} while, on the other hand, the semi-classical approach have shown tremendous success~\cite{lewenstein1994theory}. 
In this work we will show the underlying reason for this success in describing experimental observations, and we show where and how the quantum signatures appear in the field correlation functions. 

However, over the last few years, the quantum optical perspective on the process of HHG has experienced an Indian summer, with a variety of proposed, and some observed deviations from the semi-classical picture~\cite{cruz2024quantum}. 
So far the majority of this recurring interest was devoted to study and characterize the quantum state of the generated harmonics~\cite{stammer2022high, gorlach2020quantum, pizzi2023light, yi2024generation, tzur2024generation} or the state of the driving field~\cite{lewenstein2021generation, rivera2022strong, stammer2024entanglement}. This was done by solving the time-dependent Schrödinger equation for the quantized electromagnetic field coupled to matter, ranging from atoms~\cite{lewenstein2021generation, stammer2023quantum} to solid state systems~\cite{lange2024electron, gonoskov2024nonclassical, rivera2024nonclassical, lange2025excitonic}. 
Focusing on the quantum state of the field has shown interesting aspects of the interplay between strong laser fields and matter. It was theoretically demonstrated that correlated materials can imprint their correlations on the quantum state of the harmonics~\cite{pizzi2023light}, that the quantum state can exhibit squeezing~\cite{stammer2024entanglement, tzur2024generation}, and that post-selection schemes allow for the generation of non-classical states~\cite{lewenstein2021generation, rivera2022strong, stammer2022theory}.  

While the study of the quantum state of the field is interesting by itself, and provides insights into the light-matter interaction, the physically relevant objects originate from measurements such as expectation values and correlations. While the quantum state in principle allows to compute any expectation value, it can become challenging when practically doing so, for instance when computing two-time averages. 
In contrast, this work considers the complementary perspective, namely the perspective of an observer. With this, we ultimately go beyond the quantum state of the field, as we are interested in the rich manifold of observables. 
Therefore, in this work, we develop a Heisenberg picture approach for the strong laser driven process of high harmonic generation which allows to investigate observations inaccessible before. 
A clear advantage of the Heisenberg approach established in this work is that we do not rely on specific assumptions underlying the strong field driven dynamics, and that approximate solutions of the Schrödinger equation are not needed. This allows to uncover so far unknown properties of the harmonic radiation.
In particular, we gain further insights into the constitution of previously considered observables and show where to find new non-classical signatures. Further, we consider novel field observables which reveal so far unseen non-classical properties in the radiation emitted in HHG. 
This is achieved by providing a description of the first and second order field correlation function. With this, we can precisely reveal the classical and quantum nature of the spectrum and highlight on the importance of the many atom picture for the success of the semi-classical approach to HHG. 
Our analysis allows to study time dependent properties of the field from a quantum optical perspective, and particularly enables to compute two-time averages of the scattered field. Finally, we present the first theoretical prediction of photon anti-bunching in HHG. 
A schematic configuration of experimental realizations we are concerned with---field and intensity correlation functions---are presented in Fig~\ref{fig:intro}.

If we wish to summarize this work in other words, we \textit{establish and formulate a full quantum optical theory of field correlations and coherence for the process of high harmonic generation.}

The remaining article is organized as follows:
In Sec.~\ref{sec:heisenberg_picture}, we introduce the Heisenberg picture approach to HHG. Followed by Sec.~\ref{sec:1}, where we provide the quantum optical theory of field correlation functions for high harmonic generation, with the derivation of the first order field correlations in Sec.~\ref{sec:1st_order}, and show in Sec.~\ref{sec:coherence_1st_order} that HHG has first order coherence. In Sec.~\ref{sec:comparison_schrödinger} we show the connection to the previous quantum optical approaches to HHG, and go beyond those results in Sec.~\ref{sec:1st_order_incoherent}. We compare the single with the many atom picture in~\ref{sec:many_N}.  
Next, Sec.~\ref{sec:coherence_2nd_order} discuss the second order field correlation function. In Sec.~\ref{sec:intensity_correlation} we consider intensity correlations in HHG, and in Sec.~\ref{sec:manyN_2nd_order} we likewise discuss the many atom picture together with the importance for observing non-classical signatures experimentally.
In Sec.~\ref{sec:conclusion}, we conclude with the results of this article and provide a detailed outlook for future studies in Sec.~\ref{sec:outlook}.

\section{\label{sec:heisenberg_picture}QO-HHG in the Heisenberg picture}

In this work we are interested in formulating a full quantum optical theory of field correlation functions for the process of high harmonic generation (HHG), and to develop the corresponding quantum theory of optical coherence. 
Previous studies on the quantum optical description of HHG have focused on finding the final field state after the interaction with the material system, from which time-independent expectation values of field operators are obtained. However, a substantial part in understanding the radiation properties is missing, which includes information about two-time averages and operator products encoded in the field correlation functions. Furthermore, there is yet no existing quantum theory of optical coherence for the process of harmonic generation.

We are therefore interested in providing an alternative quantum optical description for the process of HHG. This is achieved by developing a Heisenberg picture approach, such that two point field expectation values can be obtained straightforwardly, and allows to study the phenomena of correlation functions of the radiation in HHG.
To provide the quantum optical theory of HHG in the Heisenberg picture we shall first define the Hamiltonians governing the dynamics of the optical field interacting with the HHG medium. The total Hamiltonian of the field plus the HHG medium is given by
\begin{align}
\label{eq:hamiltonian_total}
    H & = H_F + H_S + H_{I}, \\
    H_F & = \sum_{\lambda=1}^2 \sum_{q=1}^{q_c} \hbar \omega_{q,\lambda} a_{q, \lambda}^\dagger a_{q,\lambda}, 
\end{align}
where $H_F$ and $H_S$ are the free Hamiltonians of the field and the matter system, respectively. For the field modes we consider the driving field mode $q=1$ until the HHG cut-off at $q_c$. The sum over $\lambda$ accounts for the two possible polarization directions of the field. The system Hamiltonian $H_S$ of the HHG medium consists of $N$ charges, e.g. electrons, which are driven by the intense laser field. To allow for arbitrary matter systems, and for the sake of generality, we consider a generic system Hamiltonian $H_S$, which will be specified for concrete cases later on. This could, for instance, be a single active electron from a gas phase atom or a complex solid state system with non-trivial characteristics.
The light-matter interaction Hamiltonian $H_{I}$ is given within the dipole approximation and length-gauge where each electron $(i)$ is coupled to the electric field operator $\vb{E}_Q$ via its dipole moment operator $\vb{d}_i$ (for a detailed derivation of the interaction Hamiltonian see Ref.~\cite{stammer2023quantum})
\begin{align}
\label{eq:hamiltonian_interaction}
    H_{I} & = - \sum_{i=1}^N\vb{d}_i \cdot \vb{E}_Q, 
\end{align}
and the electric field operator is given by 
\begin{align}
    \opE = - i \tilde g \sum_{q, \lambda} \sqrt{q} \pol_{q, \lambda} (a_{q, \lambda}^\dagger - a_{q, \lambda}),
\end{align}
where $\tilde g \equiv \sqrt{\hbar \omega/(2V \epsilon_0) }$ is the coupling constant of the light-matter interaction with the quantization volume $V$, $\pol_{q, \lambda}$ is the unit polarization vector of harmonic mode $q$ with polarization $\lambda$ and $a_{q, \lambda}^{(\dagger)}$ is the annihilation (creation) operator of the respective field mode. 
The initial boundary conditions imposed on the experiment are given by a coherent state for the fundamental mode $\ket{\alpha}$ corresponding to a classical driving laser field, while the harmonic modes $q\ge 2$ are in the vacuum $\ket{0_q}$. The HHG medium is initially in the ground state of the system Hamiltonian $H_S$ such that $H_S \ket{\vb{\bar g}} = E_{\vb{g}} \ket{\vb{\bar g}}$ is the eigenstate of lowest energy $E_{\vb{g}}$. The total initial state is therefore given by
\begin{align}
\label{eq:initial_state}
    \ket{\Psi(0)} = \ket{\alpha} \otimes \ket{\{ 0_q \}} \otimes \ket{\vb{\bar g}},
\end{align}
where $\ket{ \{ 0_q \} } = \bigotimes_{q\ge 2} \ket{0_q}$ is the initial state of all harmonics. 

Now, our aim is to compute field observables, such as two-time averages. In the Heisenberg picture the operators acting on the field are time-dependent $\mathcal{O}(t,t^\prime)$, and the expectation values are evaluated over the initial state 
\begin{align}
\label{eq:expval_generic}
    \expval{\mathcal{O}(t,t^\prime)} = \operatorname{Tr}[\mathcal{O}(t,t^\prime) \rho(0)],
\end{align}
where $\rho(0)$ is the initial state of the total system (for the pure state in Eq.~\eqref{eq:initial_state} it is given by $\rho(0) = \dyad{\Psi(0)}$), and the trace is performed over matter and field degrees of freedom. 
The time-dependent field operator in the Heisenberg picture is given by 
\begin{align}
\label{eq:aq_heisenberg_definition}
    a_{q, \lambda}(t) = U^\dagger (t) a_{q, \lambda}(0) U(t),
\end{align}
where $U(t)$ is the time-evolution operator governed by the total Hamiltonian. 
To obtain the Heisenberg equation of motion (EOM), we differentiate \eqref{eq:aq_heisenberg_definition} and find (see Appendix \ref{app:heisenberg_field} for a detailed derivation)
\begin{align}
\label{eq:aq_heisenberg_eom}
    \dv{t} a_{q, \lambda}(t) = - i \omega_{q, \lambda} a_{q, \lambda}(t) + g \sqrt{q} \sum_{i=1}^N \pol_{q, \lambda} \cdot \vb{d}_i(t),
\end{align}
where $\vb{d}_i(t) = U^\dagger(t) \vb{d}_i(0) U(t) $ is the dipole moment operator in the Heisenberg picture and $g \equiv \tilde g / \hbar = \sqrt{\omega/(2 V \hbar \epsilon_0)}$.
Solving the EOM of the field operator we find 
\begin{align}
\label{eq:aq_heisenberg_solution}
    a_{q, \lambda}(t) = &  a_{q, \lambda}  e^{- i \omega_{q, \lambda} t} \\
    & + g \sqrt{q} \int_{t_0}^t \dd t^\prime e^{- i \omega_{q, \lambda} (t-t^\prime)} \sum_{i=1}^N \pol_{q, \lambda} \cdot \vb{d}_i(t^\prime). \nonumber
\end{align}

The first term in the expression is the evolution of the free field, without the presence of any charge distribution, while the second term is the contribution to the scattered field due to the presence of a time-dependent dipole moment. 
Before proceeding with the derivation, we shall emphasize on an interesting aspect of the source term in Eq.~\eqref{eq:aq_heisenberg_solution}, which concerns the temporal integral over the time-dependent dipole moment. It shows that the field at time $t$ does not only depend on the source at this particular time $t$, but depends on the entire history of the dipole from $t_0$ to $t$, emphasizing non-Markovian memory effects in the dynamics~\cite{wodkiewicz1976markovian, ishkhanyan2021markoff}.

In order to compute expectation values for the field we need an expression for the dipole moment operator in the Heisenberg picture $\vb{d}(t)$, since the trace in Eq.~\eqref{eq:expval_generic} is performed over the matter and field degrees of freedom. However, the exact solution of the Heisenberg EOM for the dipole moment is involved~\cite{sundaram1990high, diestler2008harmonic}, and does not allow to gain intuitive insights into the dynamics. This is due to the backactions between the light and matter onto each other. The light field induces a charge dynamics which in turn changes the field, which again influences the charge dynamics, and consequently approximations are needed in this hierarchy. 
In any case, the presence of the intense driving field allows to perform approximations based on physical arguments such that an expression for the time-dependent dipole moment can be obtained. 

For the EOM of the dipole moment operator it is important to note that the charge distribution is driven by an intense classical light field, which can be included in the Hamiltonian by means of the classical electric field $\vb{E}_{cl}(t)$ coupled to the dipole moment operator (for a derivation of this classical interaction Hamiltonian we refer to Appendix \ref{app:classical_interaction})
\begin{align}
\label{eq:interaction_semiclassical}
    H_{I,cl} (t) = - \sum_{i=1}^N \vb{d}_i \cdot \vb{E}_{cl}(t),
\end{align}
where the classical electric field is given by 
\begin{align}
    \vb{E}_{cl}(t) = \operatorname{Tr}[\vb{E}_Q(t) \rho(0)] = \mathbf{E}_{1, \lambda} \sin[\omega t - \varphi],
\end{align}
where $\mathbf{E}_{1, \lambda} = 2 \hbar g \abs{\alpha} \pol_{1, \lambda}$ is the classical electric field amplitude and $\varphi = \arg(\alpha)$ is the phase of the coherent state amplitude $\alpha = \abs{\alpha} e^{i \varphi}$ of the driving laser.
In analogy to the Heisenberg EOM for the field operator, the differential equation for the dipole moment is given by 
\begin{align}
\label{eq:heisenberg_dipole}
    \dv{t} \vb{d}_i(t) = \frac{i}{\hbar} U^\dagger(t) [H(t), \vb{d}_i ] U(t),
\end{align}
where now $H(t)$ is the total Hamiltonian including the semi-classical interaction $H_{I,cl}(t)$. 

To solve the EOM we need to solve the commutator with the total Hamiltonian, which turns out to be a difficult task due to the aforementioned backaction of the light field and charge. However, owing to the fact that the classical light field is of high intensity we can neglect the contribution of the backaction, and only consider the semi-classical interaction of the classical field with the dipole moment \cite{diestler2008harmonic}. Thus, the Hamiltonian in the commutator of Eq. \eqref{eq:heisenberg_dipole} is approximated by the semi-classical Hamiltonian
\begin{align}
    H_{sc} (t) = H_S + H_{I,cl}(t) = H_S - \sum_{i=1}^N \vb{d}_i \cdot \vb{E}_{cl}(t).
\end{align}

This semi-classical Hamiltonian is the conventional one used to describe the process of HHG in the semi-classical theory~\cite{lewenstein1994theory}. 
Therefore, instead of explicitly solving the exact Heisenberg EOM for the dipole moment, we use the formal solution in this approximate semi-classical frame 
\begin{align}
\label{eq:dipole_semi-classical}
    \vb{d}_i(t) = U_{sc}^\dagger (t) \vb{d}_i (0) U_{sc}(t), 
\end{align}
with the semi-classical propagator 
\begin{align}
    U_{sc}(t) = \mathcal{T} \exp[- \frac{i}{\hbar} \int_{t_0}^t dt^\prime H_{sc}(t^\prime)],
\end{align}
where $\mathcal{T}$ indicates time-ordering. 
Now, we are equipped with everything to look at the coherence properties of the harmonic radiation. We have the Heisenberg operators for the field in Eq.~\eqref{eq:aq_heisenberg_solution}, with the time-dependent dipole moment source term. We have used the intense driving field to our advantage, and consider the dipole moment to be essentially driven by the semi-classical interaction only, which encodes all the non-linearity of the HHG process in the dipole oscillations. We shall now proceed by computing the field correlation functions, allowing us to gain further insights into the distinction of the classical and quantum signatures of the field observables.

\section{\label{sec:1}Quantum theory of optical coherence in high harmonic generation}

Of special interest for characterizing the radiation field are the first-order and second-order correlation functions. For a single mode they are respectively given by 
\begin{align}
    G^{(1)}(t,t+\tau) & \propto \expval{a^\dagger(t) a(t+\tau)}, \\
    G^{(2)}(t,t+\tau) & \propto \expval{a^\dagger(t) a^\dagger (t+\tau) a(t+\tau) a(t)},
\end{align}
and are related to the spectrum, and the photon statistics, respectively. The field operators $a^{(\dagger)}(t)$ in the two-time averages are defined in the Heisenberg representation. 
The field correlation functions are given in normal ordering due to its relation to the type of measurement using photodetection, where it can be shown that every measurement using photodetection must be represented in normal ordering~\cite{mandel1995optical}.  
It is noteworthy, that the theory of photon detection and photon counting is well understood with a rich history in quantum optics~\cite{glauber1963photon, glauber1963quantum, glauber1963states, srinivas1981photon, rousseau1977new, mollow1968quantum, kelley1964theory, grochmalicki1991squeezed}, and especially the quantum theory of optical coherence by Glauber was developed based on photon detection and field correlation functions~\cite{glauber1963photon, glauber1963quantum, glauber1963states}. Note that in parallel E.~C.~G.~Sudarshan developed a quantum theoretic approach to optical coherence based on phase-space distributions~\cite{sudarshan1963equivalence}. 

We further emphasize that photon detection theory is particularly important for measuring the power spectrum of a light field, which can be shown to be strongly related to the auto-correlation function of the underlying process~\cite{mandel1995optical}. For stationary random processes, the auto-correlation function of the process and its power spectral density form a Fourier transform pair, and is known as the Wiener-Khintchine theorem~\cite{wiener1930generalized, khintchine1934korrelationstheorie}. 
However, not all light fields are stationary such that a time-dependent spectrum, introduced by Eberly and Wódkiewicz, is the proper "physical spectrum"~\cite{eberly1977time}. Furthermore, it is inherently related to the filter bandwidth used in the specific experimental implementation to measure the spectrum. 

Therefore, for the transient processes considered in this work, there are two approaches to characterize the spectrum of the field: (i) we measure all photons emitted at some frequency $\omega$, given a narrow spectral filter and collecting over a long enough temporal average ("quasi stationary case"), or (ii) we have a broader filter and measure the time-dependent spectrum of light~\cite{eberly1977time, eberly1980time}. 
Note that the time-dependent physical spectrum by Eberly and Wódkiewicz is equivalent to the power spectrum based on the Wiener-Khintchine theorem in the case of an ideal (zero bandwidth) spectral filter for stationary fields~\cite{eberly1977time}.

Together with the quantum optical description of HHG in the Heisenberg picture from Sec.~\ref{sec:heisenberg_picture}, we can compute the first order correlation function and hence the HHG spectrum. We will discuss the different contributions to the scattered field, i.e. the coherent and incoherent emission, as well as the role of the single, few and many atom picture. This provides further insights on why semi-classical approaches have been successful for decades, and how non-classical signatures appear in the observations.

\subsection{\label{sec:1st_order}First order correlations and the HHG spectrum}

To obtain the first order field correlation function, we shall relate the scattered field to the source of the radiation, such that the correlation functions can be expressed in terms of atomic operators. Then, we can solve the atomic dynamics driven by a classical intense field with established methods from strongly laser driven systems~\cite{amini2019symphony, smirnova2014multielectron}. 

To compute field expectation values we first decompose the electric field operator in the Heisenberg picture into, $\opE (t) = \vb{E}^{(+)}(t) + \vb{E}^{(-)}(t) $, with a positive and negative field component 
\begin{align}
    \vb{E}^{(+)}(t) & =  i \hbar g \sum_{q, \lambda} \sqrt{q} \pol_{q, \lambda} a_{q, \lambda}(t), \\
    \vb{E}^{(-)}(t) & = \left[\vb{E}^{(+)}(t) \right]^\dagger,
\end{align}
with the Heisenberg operator $a_q(t)$ from Eq.~\eqref{eq:aq_heisenberg_solution}, which allows to write the field operator as 
\begin{align}
    \vb{E}^{(+)}(t) = \vb{E}_f^{(+)}(t) + \vb{E}_s^{(+)}(t),
\end{align}
with 
\begin{align}
    \vb{E}_f^{(+)}(t) = i \hbar g \sum_{q, \lambda} \sqrt{q} \pol_{q, \lambda} a_{q, \lambda} e^{- i \omega_{q, \lambda} t},
\end{align}
and
\begin{align}
\label{eq:scattered_field}
    \vb{E}_s^{(+)}(t) &=  i \hbar g^2 \sum_{q, \lambda} q \pol_{q, \lambda} \int_{t_0}^t \dd t^\prime e^{- i \omega_{q, \lambda} (t-t^\prime) } \\
    &\quad \times \sum_{i=1}^N \pol_{q, \lambda} \cdot \vb{d}_i (t^\prime).  \nonumber
\end{align}

Here, $\vb{E}_f^{(+)}(t)$ describes the evolution of the free field in the absence of sources, and $\vb{E}_s^{(+)}(t)$ is the source field radiated by the atomic scatterer. 
From now on, we implicitly assume that the driving field is linearly polarized such that only one polarization component contributes to the dynamics, and we therefore omit the polarization index $\lambda$ in the remainder of the manuscript. Yet, we note that the theory allows to consider arbitrary polarization configurations of the fields (for a recent quantum optical approach using structured light we refer to Ref.~\cite{rivera2024non}).

We can now compute normal-ordered correlation functions, such as field 
\begin{align}
\label{eq:first_order_coherence}
    G^{(1)} (t_1, t_2) \equiv \expval*{E^{(-)}(t_1) E^{(+)}(t_2) },
\end{align}
and intensity correlations
\begin{align}
\label{eq:second_order_coherence}
    G^{(2)}(t_1,t_2) \equiv \expval*{E^{(-)}(t_1)E^{(-)}(t_2)E^{(+)}(t_2)E^{(+)}(t_1)}.
\end{align}

\begin{figure}
    \centering
	\includegraphics[width = 1\columnwidth]{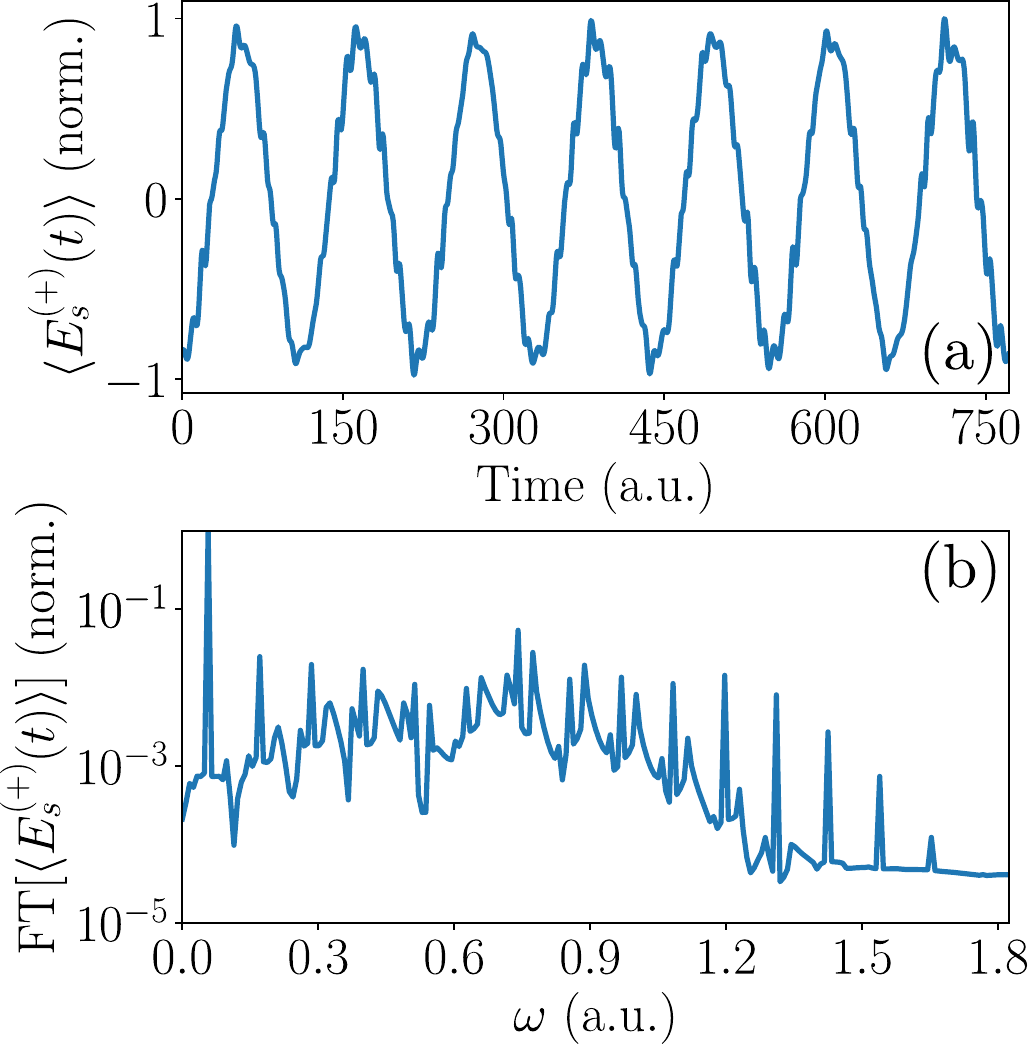}
	\caption{In (a) we show the expectation value of the total scattered field $\expval*{\vb{E}_s^{(+)}(t)} \equiv \sum_q \expval*{\vb{E}_{s,q}^{(+)}(t)}$, and in (b) the Fourier transform (FT) of the scattered field from (a). The FT of the field in (a) shows the typical characteristics of an HHG spectrum. The field parameters were chosen as $E_0 = 0.053$ a.u. and $\omega_L = 0.057$ a.u., while for the atomic ionization potential $I_p = 0.5$ a.u., corresponding to that of hydrogen.}
      \label{fig:scattred_field}
\end{figure}

However, first, we want to consider the mean value of the scattered field. Since the harmonics are initially in the vacuum, the only contribution to the field is given by the source term
\begin{align}
    \lim_{t \to \infty} \expval*{\vb{E}_{s,q}^{(+)}(t)} = i \hbar g^2 q \pol_q \sum_{i=1}^N \pol_q \cdot \expval{\vb{d}_i(\omega_q)} e^{- i \omega_q t},  
\end{align}
where we have only considered the scattered field of the $q$-th harmonic and $\expval{\vb{d}_i (\omega_q)} = \int_{- \infty}^\infty \dd t \expval{\vb{d}_i(t)} e^{i \omega_q t}$ is the Fourier component of the time-dependent dipole moment expectation value.
As expected, it shows that the field components oscillate at the harmonic frequencies $\omega_q$, in agreement with the established semi-classical result for classical coherent harmonic radiation~\cite{lewenstein1994theory}. The dynamical picture is given by coherent emission from the induced dipole moment, and the high-order components are given by the non-linear oscillations of the dipole. In comparison with the classical and the semi-classical picture, which are given by a dipole oscillator or the dipole moment expectation value, respectively, this quantum optical result is in full agreement and represents classical coherent radiation.
In Fig.~\ref{fig:scattred_field}~(a) we present the expectation value of the scattered field summed over all harmonic frequencies, $\expval*{\vb{E}_s^{(+)}(t)} \equiv \sum_q \expval*{\vb{E}_{s,q}^{(+)}(t)}$. It shows the expected dominant oscillation on the fundamental driving frequency, in addition to the anharmonic contributions to the oscillations of the scattered field. The corresponding classical spectrum is obtained from the Fourier transform of the time-dependent field, and shown in Fig.~\ref{fig:scattred_field}~(b), with the well known structure of a generic HHG spectrum. It includes the HHG plateau as well as the typical cut-off.

Nevertheless, this essentially (semi)-classical picture is not complete. Due to the quantum mechanical nature of the dipole moment operator we have a probabilistic process, and the oscillator amplitude can in principle acquire a stochastic component. Therefore, in general, the HHG spectrum should not be calculated from the mean value of the scattered field, but from the Fourier transform of the auto-correlation function in Eq.~\eqref{eq:first_order_coherence}. 
Hence, we consider the first order correlation function of one harmonic mode 
\begin{align}
\label{eq:G1_definition_harmonic}
    G^{(1)} (t, t+\tau) = \hbar^2 g^2 q \expval{a_q^\dagger(t) a_q(t+\tau)}. 
\end{align}

Using the field correlation function, we shall now compute the proper HHG spectrum, and study the coherence properties of the harmonic radiation. 
Using Eq. \eqref{eq:aq_heisenberg_solution} we have 
\begin{align}
\label{eq:first_order_a}
    \langle a_q^\dagger &   (t) a_q(t+\tau) \rangle  = g^2 q e^{- i \omega_q \tau} \int_{t_0}^t dt_1 \int_{t_0}^{t+\tau} dt_2 e^{- i \omega_q (t_1 - t_2)} \nonumber \\
    & \times \sum_{i,j = 1}^N \bra{\vb{\bar g}}[\pol_q \cdot \vb{d}_i (t_1)] [\pol_q \cdot \vb{d}_j (t_2)] \ket{\vb{\bar g}}, 
\end{align}
where we have used that the field of the harmonic modes $q \ge 2$ are initially in the vacuum, and therefore the expectation value of the free evolution vanishes. 
We can see that the first order correlation function of the harmonic radiation, and consequently also the HHG spectrum, is determined by the dipole moment correlations. This involves dipole correlations between different emitters $i$ and $j$, as well as correlations at different times from the same emitter. Since the main goal of this work is to introduce the concepts and methods of the quantum optical approach to optical coherence for HHG by building the general theory, we mainly focus on atomic gas targets. All notions of field correlations functions developed in this work can later be used for all other target systems.

To get an idea about the first order correlation function, we consider the simplest scenario of uncorrelated atoms such that the dipole moments of different electrons are independent. 
We further consider the single atom picture $N=1$, which, however, is justified due to the agreement with semi-classical perspective of HHG~\cite{sundaram1990high}, and it will become clear in Sec.~\ref{sec:many_N} how the many atom case differs. 
Therefore, we have for the expectation value of the matter part in Eq.~\eqref{eq:first_order_a}, along the laser polarization direction, $\pol_q \cdot \vb{d}_i(t) \equiv d_i(t)$; 
\begin{align}
\label{eq:dipole_single_atom}
    \expval{d(t_1) d(t_2)} = \expval{d(t_1)} \expval{d(t_2)} + \expval{\Delta d(t_1) \Delta d(t_2)},
\end{align}
where we have defined the fluctuations of the dipole around its mean $\Delta d(t) \equiv d(t) - \expval{d(t)}$.
Consequently, the first order correlation function can be decomposed into two terms 
\begin{align}
\label{eq:G1_longtime}
    \lim_{t \to \infty} G^{(1)} (t,t+\tau) \equiv G^{(1)}_{coh}(\tau) + G^{(1)}_{inc}(\tau),
\end{align}
where we have considered the long time limit as $t \to \infty$, and we have defined 
\begin{align}
\label{eq:Gone_coherent}
    \Gone_{coh} (\tau) & \equiv \hbar^2 g^4 q^2 e^{- i \omega_q \tau} \abs{\int_{-\infty}^\infty dt e^{- i \omega_q t } \expval{d(t)} }^2 \\
    & = \hbar^2 g^4 q^2 e^{- i \omega_q \tau} \abs{\expval{d(\omega_q)}}^2,
\end{align}
and 
\begin{align}
\label{eq:Gone_incoherent}
    \Gone_{inc}(\tau)  & \equiv   \hbar^2 g^4 q^2  e^{- i \omega_q \tau} \int_{-\infty}^\infty dt_1 \int_{-\infty}^\infty dt_2 e^{- i \omega_q (t_1 - t_2)} \nonumber \\
    & \times \expval{\Delta d(t_1) \Delta d(t_2)},
\end{align}
which are the coherent and incoherent contribution to the first order correlation function, respectively. We have further assumed that the initial time $t_0 \to \infty$ is way before the pulse starts.
We can see that the coherent contribution to the first order correlation is determined by the Fourier transform of the time-dependent dipole moment expectation value. This is akin to the HHG spectrum on the semi-classical description, and we shall see below its fundamental connection. 
Furthermore, we see that in addition to the coherent contribution of the dipole expectation value, there is an incoherent contribution from the correlations of the two-time dipole moment fluctuations around the mean.
While the coherent contribution to the field gives essentially the same result as a classical dipole undergoing non-linear oscillations, or like established semi-classical models, it is expected that the incoherent part leads to deviations from the classical picture. 
This is because the typical classical and semi-classical models do not go beyond mean field descriptions, while the incoherent part due to the dipole fluctuations takes into account the full probabilistic quantum nature of the process. 

We are now ready to compute the HHG spectrum using the first order correlation function~\cite{mandel1995optical, eberly1992spectrum, cresser1983theory}. 
Therefore, we will make use of the Wiener-Khintchine theorem for fluctuating processes~\cite{wiener1930generalized, khintchine1934korrelationstheorie}, stating that for stationary random processes the power spectral density $S(\omega)$ and the first order auto-correlation function $\Gone (t,t+\tau)$ are a Fourier transform pair.
Consequently, we consider the long time limit of the auto-correlation function $\lim_{t \to \infty} \Gone (t,t+\tau) \equiv \Gone(\tau)$, and assume that the emitted field becomes quasi-stationary in the sense that the two-time average only depends on the time-difference. This is justified when the driving pulse consists of enough field cycles, such that the time of averaging does not depend on the temporal location within the pulse. The Wiener-Khintchine theorem, together with the results from Eq.~\eqref{eq:G1_longtime}, can then be used to compute the corresponding HHG spectrum $S(\omega)$, and is given by 
\begin{align}
    S(\omega) & = \frac{1}{\pi} \operatorname{Re}\left[ \int_0^\infty d\tau \lim_{t \to \infty} \Gone (t,t+\tau) e^{i \omega \tau} \right] \\
    & = S_{coh}(\omega) + S_{inc}(\omega),
\end{align}
which alike is decomposed into a coherent and an incoherent contribution. 
We shall first look at the coherent part of the power spectrum and find that 
\begin{align}
\label{eq:spectrum_coherent}
    S_{coh}(\omega) & =  \frac{\hbar^2 g^4 q^2}{\pi} \abs{\expval{d(\omega_q)}}^2 \operatorname{Re} \left[ \int_0^\infty d\tau e^{i (\omega - \omega_q)\tau} \right] \\
    & = \hbar^2 g^4 q^2 \abs{\expval{d(\omega_q)}}^2 \delta (\omega - \omega_q).
\end{align}

\begin{figure*}
    \centering
	\includegraphics[width = \textwidth]{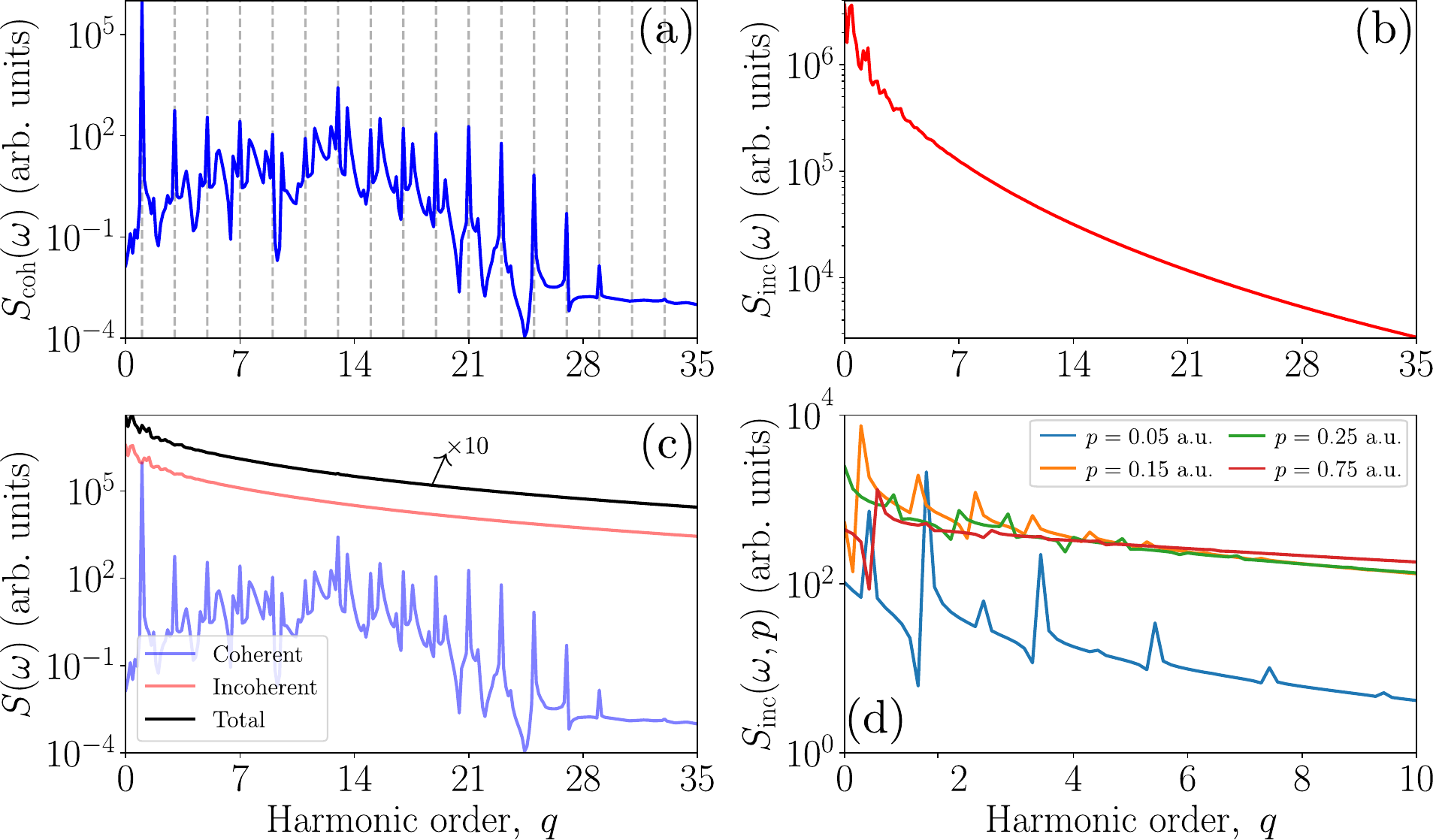}
	\caption{Here we show in (a) the coherent spectrum $S_{coh}(\omega)$ from Eq.~\eqref{eq:spectrum_coherent}, in (b) the incoherent spectrum $S_{inc}(\omega)$ from Eq.~\eqref{eq:spectrum_incoherent}. In (c) the total spectrum of both contributions are summed (with the individual contributions in faded colors). In (d) we show the incoherent part of the spectrum $S_{inc}(\omega)$ for a few specific electron canonical momenta $p$. The field parameters are the same as those in Fig.~\ref{fig:scattred_field}.}
      \label{fig:spectrum_comparison}
\end{figure*}

It shows that the HHG spectrum consists of peaks at frequencies $\omega_q = q \omega$, with a weight given by the Fourier transform of the time dependent dipole moment expectation value, while the dipole moment takes into account the symmetry of the process leading to the observation of only odd harmonics due to conservation of parity and energy~\cite{perry1993high}. 
In Fig.~\ref{fig:spectrum_comparison}~(a) we show the coherent contribution to the spectrum from Eq.~\eqref{eq:spectrum_coherent}, exhibiting the well known plateau and cut-off structure as well as peaks at odd multiple of the driving frequency. 
The width of the peaks in the harmonic spectrum is determined by the Fourier components of the dipole moment expectation value, in which the envelope of the driving field can be taken into account to mimic the bandwidth of the driving field~\cite{lewenstein2021generation, stammer2023quantum}.

We shall make a few comments on how we obtained the HHG spectrum. In Eq.~\eqref{eq:G1_longtime} we have considered the long-time limit of the auto-correlation function, such that we can consider the light field obtained from the dipole source as "quasi-stationary", in the sense that for long enough driving pulses the emission does not change. Consequently, the auto-correlation function, as given in Eq.~\eqref{eq:Gone_coherent} and Eq.~\eqref{eq:Gone_incoherent}, only depends on the time difference $\tau$. Further, in Eq.~\eqref{eq:G1_definition_harmonic} we have considered the correlation function $G^{(1)}(t,t+\tau)$ only for a given harmonic mode $q$, which corresponds to the implicit assumption that the spectral measurement includes a narrow bandwidth filter (in the ideal case infinitesimally small, corresponding to a delta function). 
The "quasi-stationary" limit, together with the narrow bandwidth filter, allows to use of the Wiener-Khintchine theorem for calculating the HHG spectrum since it is equivalent to the time-dependent spectrum by Eberly and Wódkiewicz in this case~\cite{eberly1977time}.

\subsection{\label{sec:coherence_1st_order}First order coherence in HHG}

Having computed the first order field correlation function in the previous section, we shall now look at the first order coherence property of the generated harmonics. 
For the first order correlation function it holds that 
\begin{align}
    \abs{G^{(1)}(t_1,t_2)}^2 \leq \abs{G^{(1)}(t_1,t_1) G^{(1)}(t_2,t_2) }, 
\end{align}
and a field is said to possesses first order optical coherence when the equality holds. Defining the normalized correlation function
\begin{align}
    g^{(1)}(t_1,t_2) \equiv \frac{G^{(1)}(t_1,t_2)}{\abs{G^{(1)}(t_1,t_1) G^{(1)}(t_2,t_2)}^{1/2}},
\end{align}
the condition of first order optical coherence is given by 
\begin{align}
    \abs{g^{(1)}(t_1,t_2)} = 1. 
\end{align}

Now, for the correlation function of the harmonics in Eq.~\eqref{eq:G1_definition_harmonic}, we have to compute 
\begin{align}
    g^{(1)}(t_1,t_2) = \frac{ \expval*{a^\dagger_q(t_1) a_q(t_2)} }{ \abs{ \expval*{a^\dagger_q(t_1) a_q(t_1)} \expval*{a^\dagger_q(t_2) a_q(t_2)} }^{1/2} },
\end{align}
which boils down to consider dipole moment expectation values. Considering $t_1 = t$, and $t_2 = t+\tau$ as well as the long time limit from Eq.~\eqref{eq:G1_longtime}, we have for the normalized first order correlation function
\begin{align}
    g^{(1)}(\tau) = \frac{ G^{(1)}_{coh}(\tau) + G^{(1)}_{inc}(\tau) }{ \abs{G^{(1)}_{coh}(0) + G^{(1)}_{inc}(0)} }.
\end{align}

Using that $G^{(1)}(\tau) = e^{- i \omega_q \tau} G^{(1)}(0)$ in the long time limit, as we can see from Eq.~\eqref{eq:Gone_coherent} and \eqref{eq:Gone_incoherent}, we find 
\begin{align}
    \abs{\lim_{t \to \infty} g^{(1)}(t,t+\tau)} = 1,
\end{align}
showing that HHG possess first order optical coherence, and implies that harmonic radiation shows Young type interference in which the amplitudes of two fields are superposed and interfere~\cite{nabekawa2006interferometric, nabekawa2009interferometry}.

\subsection{\label{sec:comparison_schrödinger}Connection to previous approaches}

In this section, we shall make a little detour to comment on an interesting observation with respect to the previous quantum optical approaches to HHG~\cite{lewenstein2021generation, rivera2022strong, gorlach2023high}. Our observation for the coherent contribution to the spectrum is in agreement with the previous quantum optical theory of HHG within the Schrödinger picture, in which an approximate solution of the time-dependent state of the field was derived~\cite{lewenstein2021generation, rivera2022strong}. This is obtained in the limit of negligible ground state population of the electron, corresponding to vanishing dipole moment correlations~\cite{sundaram1990high}. 
In this picture the approximate Hamiltonian of the light-matter interaction, after unitary transformation into a suitable frame~\cite{lewenstein2021generation}, is given by
\begin{align}
\label{eq:hamiltonian_approximate}
    H_{app.}(t) = - \expval{\vb{d}(t)} \cdot \vb{E}_Q(t),  
\end{align}
where $\expval{\vb{d}(t)} = \bra{\text{g}} \vb{d}(t) \ket{\text{g}}$ is the time-dependent dipole moment expectation value evaluated in the electronic ground state of a single atom.
This simplified interaction Hamiltonian has the crucial difference to Eq.~\eqref{eq:hamiltonian_interaction}, that instead of the dipole moment operator, the expectation value of the dipole moment is coupled to the field. This implies that the final field state propagated by this Hamiltonian $H_{app.}(t)$ can be obtained analytically exact, giving rise to an expression linear in the field operators. The corresponding solution for the same initial state as in Eq.~\eqref{eq:initial_state} is given by a product of coherent states 
\begin{align}
    \ket{\Phi} = \ket{\alpha + \chi_1} \bigotimes_{q=2}^{q_c} \ket{\chi_q},
\end{align}
where the coherent state amplitudes are given by 
\begin{align}
    \chi_q = g \sqrt{q} \int_{-\infty}^\infty dt \expval{d(t)} e^{- i \omega_q t}.
\end{align}

The underlying assumption to obtain this expression are negligible dipole moment fluctuations, and we can see that only the coherent contribution to the spectrum is reproduced, which corresponds to vanishing dipole moment correlations at different times~\cite{stammer2022theory, stammer2024entanglement}. 
The observation that under these approximations the result is given by a product of coherent states~\cite{lewenstein2021generation}, there is a widely implemented approximation in quantum optical HHG, namely that coherent input states (coherent drive and vacuum for the harmonics), are mapped to coherent output states~\cite{lewenstein2021generation}. Even in the case of non-classical driving fields~\cite{gorlach2023high, stammer2024absence, rivera2024non, tzur2024generation} this approximation is used for the derivation of the final field state.  

Since the presence of these dipole moment correlations is directly linked to the population transfer during the electron dynamics, it was shown that higher order contributions beyond the mean field evolution arise if the depletion of the electronic ground state is taken into account~\cite{stammer2024entanglement}. On the level of the quantum state, this leads to entanglement and squeezing across all modes~\cite{stammer2024entanglement}.
We note that everything that comes now goes beyond this assumption of vanishing dipole fluctuations, and explicitly includes the dipole moment correlations, such that we are independent of the approximation and the underlying assumptions in these derivations.

\subsection{\label{sec:1st_order_incoherent}Including dipole moment correlations}

In Sec.~\ref{sec:1st_order} we have discussed the coherent contribution to the HHG spectrum, originating from the dipole moment expectation value representing a classical charge current. We have shown that the coherent contribution is the same as in the semi-classical model, and corresponds to classical coherent radiation. The previous section~\ref{sec:comparison_schrödinger} has shown the relation to the quantum state of the field, showing that the result is essentially classical by means of product coherent states. The results of this section go beyond the regime in which coherent input states map to coherent output states, as often assumed in the quantum optics approach to HHG~\cite{lewenstein2021generation, gorlach2023high}.

Now, we shall go beyond the classical regime and consider the incoherent contribution to the field arising from the dipole moment fluctuations. Evaluating the two-time dipole moment fluctuations in Eq.~\eqref{eq:dipole_single_atom}, we find 
\begin{align}
    \expval{\Delta d(t_1) \Delta d(t_2)} = \int \dd v \bra{\text{g}} d(t_1) \ket{v} \bra{v} d(t_2) \ket{\text{g}},
\end{align}
where we have used the resolution of the identity for the electron $\mathds{1} = \dyad{\text{g}} + \int \dd v \dyad{v}$. Note that we have neglected bound excited states of the electron since those can normally be neglected in the process of HHG, although the presence of a cavity~\cite{yi2024generation} or an initial superposition state can change the situation~\cite{rivera2024squeezed}, and excited electronic states can become relevant.
This can now be used to compute the incoherent contribution to the first order correlation function  
\begin{align}
    \Gone_{inc}(\tau)  \equiv & \,  \hbar^2 g^4 q^2  e^{- i \omega_q \tau} \int \dd v \, d_{\text{g}v}(\omega_q) d^*_{v\text{g}}(\omega_q),
\end{align}
which is given by the Fourier transform of the time-dependent transition matrix elements between the ground and continuum states
\begin{align}
    d_{gv}(\omega_q) \equiv \int_{-\infty}^\infty \dd t\ d_{\text{g}v}(t) e^{- i \omega_q t},
\end{align}
where we have defined $d_{\text{g}v}(t) = \bra{\text{g}} d(t) \ket{v}$.
Finally, the incoherent contribution to the spectrum reads 
\begin{align}
\label{eq:spectrum_incoherent}
    S_{inc}(\omega) = \hbar^2 g^4 q^2 \delta(\omega - \omega_q ) \int \dd v \, d_{\text{g}v}(\omega_q) d^*_{v\text{g}}(\omega_q) .
\end{align}

We can now show the incoherent contribution plus the total spectrum in Fig.~\ref{fig:spectrum_comparison}~(b) and (c), respectively. While the coherent contribution in (a) shows the signatures of a typical HHG spectrum, there is a striking difference to the incoherent part. In contrast to the coherent emission, the incoherent spectrum does not exhibit well defined peaks and is a few orders of magnitude higher. 
The absence of the peaks can be understood from the absence of the dipole moment expectation value evaluated in the ground state, which includes the selection rule of energy and parity conservation. In contrast, the incoherent contribution in Eq.~\eqref{eq:spectrum_incoherent} involves the time-dependent dipole transition matrix elements between ground and continuum states. Due to the average over all continuum states in Eq.~\eqref{eq:spectrum_incoherent} the incoherent contribution looses its peaks, which can be seen in Fig.~\ref{fig:spectrum_comparison}~(d) where we show the incoherent contribution originating from specific continuum state $\ket{v}$. The contribution from a single continuum state again shows a peaked structure for states close to the ionization threshold. 
This is in agreement with the results presented in Ref.~\cite{gorlach2020quantum}, in which electronic expectation values lead to a conventional HHG spectrum, while transitions between bound states give rise to an exponentially decaying spectrum without well defined peaks. These two contributions correspond to the dipole moment expectation value and the transitions matrix elements with continuum states in our work, respectively. We suspect that averaging over all bound states in Ref.~\cite{gorlach2020quantum} similarly yield a spectrum without peaks. 
However, it remains to discuss that apparently the total HHG spectrum exhibits no peaks (see Fig.~\ref{fig:spectrum_comparison} (c) black curve), which is due to the dominant incoherent contribution. 

Here, it is important to note that we have shown the response of a single emitter, albeit HHG experiments involve many atoms emitting harmonic radiation. 
In the next section we will show that the appearance of well defined peaks in experimental HHG spectra is due to the many atom regime, and the different scaling of the contributions with respect to the number of atoms $N$.

\subsection{\label{sec:many_N}Single atom vs. many atom regime}

In the previous sections, we have so far discussed the case of a single atom, although the general theory derived in section~\ref{sec:heisenberg_picture} allows to include $N$ different emitters. 
We can see from Eq.~\eqref{eq:first_order_a} that in the expression for the first-order field correlation function, the two-time dipole correlation function between different dipoles of the $N$ charges appears. We can evaluate the double sum and partition into two parts, such that
\begin{align}
\label{eq:dipole_product_N}
    \sum_{i,j=1}^N \bra{\vb{\bar g}} d_i(t_1) & d_j(t_2) \ket{\vb{\bar g}} =  \sum_{i=1}^N \bra{\vb{\bar g}} d_i (t_1) d_i(t_2) \ket{\vb{\bar g}} \\
    & + \sum_{i \neq j}^N \sum_{j=1}^N \bra{\vb{\bar g}} d_i (t_1) d_j(t_2) \ket{\vb{\bar g}}, \nonumber
\end{align}
where we have considered the dipole moment along the laser polarization direction, $\pol_q \cdot \vb{d}_i(t) \equiv d_i(t)$.
We have separated the double sum in such a way that the first term corresponds to $N$ independent emitters, each given by the same expression as obtained in the previous section, and the second term takes into account dipole moment correlations between the different atoms~\cite{sundaram1990high, eberly1992spectrum, xu1993non}. 
Inspecting these two terms, we find that both scale very differently with the number of atoms $N$. While the first term is of order $\mathcal{O}(N)$, the second term scales as $\mathcal{O}(N^2)$, such that in the conventional experimental setting of $N \gg 1$ the second term is much larger than the first term. Nevertheless, we proceed with the general expression and further analyze the different contributions. 

We shall first look at the second term in Eq.~\eqref{eq:dipole_product_N}, and factorize the expectation value of the operator product 
\begin{align}
\label{eq:dipole_correlations_factorize}
    \bra{\vb{\bar g}} d_i(t_1) d_j(t_2) \ket{\vb{\bar g}} \approx \bra{\text{g}_i} d_i(t_1) \ket{\text{g}_i} \bra{\text{g}_j} d_j(t_2) \ket{\text{g}_j},
\end{align}
which is of course not correct in general. 
However, factorization of the dipole moment correlations between different atoms is valid if the initial state of the system factorizes between all charges, e.g. $\ket{\vb{\bar g} } = \bigotimes_{i=1}^N \ket{\text{g}_i}$, such that the expectation values factorizes and Eq.~\eqref{eq:dipole_correlations_factorize} becomes an equality.
This is in general given in HHG experiment in gases, where the individual atoms are uncorrelated. However, in more complex targets such as solid state materials \cite{ghimire2019high, goulielmakis2022high}, systems with initial correlations \cite{silva2018high, pizzi2023light} or non-trivial topology \cite{silva2019topological, jimenez2020lightwave} this assumption about the initial state does not necessarily hold. 

Now, evaluating the double sum in Eq.~\eqref{eq:dipole_product_N} for uncorrelated emitters explicitly, we have 
\begin{align}
\label{eq:dipole_N_doublesum}
    \sum_{i \neq j}^N   \sum_{j=1}^N \bra{ \text{g}_i } d_i (t_1) \ket{\text{g}_i} \bra{\text{g}_j} d_j(t_2) \ket{\text{g}_j} \\
    = N (N-1)  \expval{d(t_1)} \expval{d(t_2)},  \nonumber 
\end{align}
where $\expval{d(t)} = \bra{\text{g}} d(t) \ket{\text{g}}$, and assuming that the expectation value is independent of the specific emitter.
This assumption, which is similar to the dipole approximation in which each atom experience the same field, the contribution of each electron is the same and we can therefore drop the subscript. 
Under the same assumptions, the first term in Eq.~\eqref{eq:dipole_product_N} can be written as
\begin{align}
\label{eq:dipole_correlation_2time}
    \sum_{i=1}^N \bra{\vb{\bar g}} d_i (t_1) d_i(t_2) \ket{\vb{\bar g}} = N \bra{\text{g}} d(t_1) d(t_2) \ket{\text{g}}. 
\end{align}

In summary, after evaluating the sum over the $N$ different emitters, we found that the incoherent contribution (Eq.~\eqref{eq:dipole_correlation_2time}) scales linear in the number of atoms, while the coherent contribution (Eq.~\eqref{eq:dipole_N_doublesum}) scales quadratic with $N$. 
Considering that usually a large number of atoms interacts with the driving laser we have $N^2 \gg N$, such that the coherent contribution to the spectrum normally dominates the incoherent contribution. It is the $N$ vs. $N^2$ scaling of the different contributions why the semi-classical picture of HHG was so tremendously successful over decades: under normal experimental situations the coherent part of the scattered field dictates the spectrum. 

It remains to discuss the role of the two-time correlations in Eq.~\eqref{eq:dipole_correlation_2time}, which also sheds light onto the Schrödinger picture approach to HHG~\cite{lewenstein2021generation, stammer2024entanglement}. It was shown that the spectrum is equivalent with the semi-classical picture in the case of, (i) vanishing dipole moment correlations at two-times, (ii) when uncorrelated atoms are assumed in order to factorize the initial state, and (iii) if a large number of emitters is considered such that the coherent part dominates due to the $N^2$ scaling. 
We note that even in the presence of temporal dipole moment correlations, the state is essentially classical due to the dominance of $N^2 \gg N$. This suggest that in this regime the final state is given by a product of coherent states as discussed in Sec.~\ref{sec:comparison_schrödinger}. 
Furthermore, in the case of negligible dipole moment correlations for a single emitter, we can improperly factorize this two-time average of Eq.~\eqref{eq:dipole_correlation_2time}, and have 
\begin{align}
    \bra{g} d(t_1) d(t_2) \ket{g} \approx \expval{d(t_1)} \expval{d(t_2)}.
\end{align}

This assumption would then ultimately lead to the following approximate expression for the total contribution of $N$ emitters 
\begin{align}
\label{eq:dipole_N_full_approximation}
    \sum_{i,j=1}^N \bra{\vb{\bar g}} d_i(t_1) & d_j(t_2) \ket{\vb{\bar g}} \approx N^2 \expval{d(t_1)} \expval{d(t_2)},
\end{align}
which is the $N^2$ scaling of the coherent contribution, and aligns with the phenomenologically introduced $N$ emitter case considered in previous work~\cite{lewenstein2021generation, rivera2022strong}. 

Finally, we can now look at the difference to the first-order correlations between the single and many atom case, and how it influences the spectrum. 
The simplest and trivial case is the one in which all assumptions and approximations leading to Eq.~\eqref{eq:dipole_N_full_approximation} are applied, which only differs from the single atom case in Eq.~\eqref{eq:Gone_coherent} by the factor of $N^2$. In this situation the single atom picture is essentially the same as the many atom picture, and is the same as the classical case~\cite{sundaram1990high}. 
The more interesting case is the study in which the dipole moment correlations are important, such that we do not perform the approximation in Eq, \eqref{eq:dipole_N_full_approximation}, and we find the following first order correlation functions
\begin{align}
    \Gone_{coh, N} (\tau) & = N^2 \Gone_{coh} (\tau), \\
    \Gone_{inc, N} (\tau) & = N \Gone_{inc} (\tau), 
\end{align}
for the coherent and incoherent contribution, respectively. 
For the incoherent contribution we have again used the re-writing of the dipole correlations alike in Eq.~\eqref{eq:dipole_single_atom}, while the first term cancels with the second term in Eq.~\eqref{eq:dipole_N_doublesum} and the dipole fluctuations remain. 
Since the many atom case has this simple relationship with the single atom case, we do not need to compute the spectrum from the auto-correlation function again.

We can conclude that in the many atom case, without correlations between the different emitters, the many atom response can essentially be obtained from the single atom picture~\cite{sundaram1990high}. They only differ by the different scaling of the coherent and incoherent contribution to the scattered field, such that in conventional experimental configurations with $N \gg 1$ the coherent contribution to the spectrum dominates the incoherent contribution $S_{coh}(\omega) \gg S_{inc}(\omega)$.
It might be this reason why the semi-classical approach of HHG was so successful over decades without the need for a full quantum optical framework when it comes to the description of the experimentally recorded spectra. 
The bottom line of this is that the coherent contribution corresponds to classical radiation, while the incoherent part is the more interesting one. However, the interesting incoherent contribution is overruled by the many atoms response participating in the process.

\section{\label{sec:coherence_2nd_order}Second order correlations }

So far, we have looked at the first order correlations of the field, which is related to the HHG spectrum. Now, we shall look at the second order correlation function
\begin{align}
\label{eq:correlation_2nd_order_general}
    G^{(2)}(t_1,t_2) \equiv \expval*{E^{(-)}(t_1)E^{(-)}(t_2)E^{(+)}(t_2)E^{(+)}(t_1)},
\end{align}
which can be re-written in terms of the normally ordered intensity correlation function 
\begin{align}
    G^{(2)}(t_1,t_2) = \expval*{: I(t_1) I(t_2) :},
\end{align}
where $: \, :$ indicates normal ordering, such that all creation operators are written on the left while all annihilation operators are on the right, and $I(t)$ is the intensity operator $I(t) = \eta n(t)$, where $\eta$ takes into account all details about the relation of intensity and photon number, and $n(t)$ is the photon number operator at time $t$.  
The second order correlation function is essentially an intensity correlation measurement, where the experiment measures the joint photocurrent of photons at time $t_1=t$ and of photons at time $t_2 = t+\tau$. Such a configuration was introduced by Hanbury Brown and Twiss (HBT), showing first signatures of photon bunching of chaotic light~\cite{brown1956correlation}. Such a HBT-configuration for an intensity correlation measurement is shown in Fig.~\ref{fig:intro}~(c).

In this section we are interested in measuring the intensity correlation of the harmonics by correlating the intensity measurements from two detectors with a time delay $\tau$. The intensity correlation for a single harmonic mode is hence given by 
\begin{align}
\label{eq:definition_intensity_correlation}
    G^{(2)}(t,t + \tau) & = \expval{:I_q(t) I_q(t+\tau):} \\
    & = \eta^2 \expval{a_q^\dagger(t) a_q^\dagger (t+\tau) a_q(t+\tau) a_q(t)},
\end{align}
for which we have defined the time-dependent intensity operator $I_q(t) \equiv \eta \, U^\dagger (t) a_q^\dagger a_q U(t) = \eta \, a_q^\dagger (t) a_q(t)$.
A dimensionless measure of such an intensity correlation is given by the normalized intensity correlation function 
\begin{align}
    g^{(2)}(\tau) & = \frac{\expval*{: I(t)I(t+\tau) :}}{\expval*{I(t)} \expval*{I(t+\tau)}} \\
    & =  \frac{\expval*{a^\dagger(t) a^\dagger(t+\tau) a(t+\tau) a(t)}}{\expval*{a^\dagger(t) a(t)} \expval*{a^\dagger(t+\tau) a(t+\tau)}}.
\end{align}

Before going into the details of the intensity correlations in HHG, we shall first consider generic statements for the variance of the classical intensity $\expval{[\Delta I(t)]^2} = \expval{[I(t) - \expval{I(t)}]^2}$, which implies that
\begin{align}
    \expval{[\Delta I(t)]^2}  = \expval{I^2(t)} - \expval{I(t)}^2 \ge 0,
\end{align}
is positive semi-definite, with the consequence that 
\begin{align}
    \expval{I^2(t)} \ge \expval{I(t)}^2.
\end{align}

We use this to provide a classical relation for the intensity correlation, which are then used to witness eventual non-classicality in the intensity correlation measurement of the harmonics from the quantum optical picture. For a classical description without a commutator structure, we have 
\begin{align}
    g_{cl}^{(2)}(\tau) = \frac{\expval{I(t) I(t+\tau)}}{\expval{I(t)} \expval{I(t+\tau)}}.
\end{align}

For vanishing time delay $\tau = 0$ we find that 
\begin{align}
    g_{cl}^{(2)} (0) = \frac{\expval{I^2(t)}}{\expval{I(t)}^2} \ge \frac{\expval{I(t)}^2}{\expval{I(t)}^2} = 1,
\end{align}
stating that for classical radiation the intensity correlation at vanishing time delay is always greater or equal to one. 
We can also look at non-vanishing time-delays for which we consider $[I(t) - I(t+\tau)]^2 \ge 0$, and use that under the assumption of a stationary signal we have $\expval{I^2(t)} = \expval{I^2(t+\tau)}$. Then, we find that $\expval{I^2(t)} \ge \expval{I(t) I(t+\tau)}$, which implies that for classical radiation 
\begin{align}
    g_{cl}^{(2)} (0) =  \frac{\expval{I^2(t)}}{\expval{I(t)}^2} \ge  \frac{\expval{I(t) I(t+\tau)}}{\expval{I(t)}^2} = g_{cl}^{(2)}(\tau),
\end{align}
showing that the intensity correlations can only decrease over time. 
With this we can summarize the classical properties of the second order correlation function 
\begin{align}
\label{eq:g2_classical1}
    g_{cl}^{(2)} (0) & \ge g_{cl}^{(2)} (\tau), \\
\label{eq:g2_classical2}
    g_{cl}^{(2)} (0) & \ge 1,
\end{align}
which originated from the fact that the classical intensity observables commute. A violation of these inequalities is a witness of non-classical properties in the corresponding measurement. The property of the normalized intensity correlation function at vanishing time-delay is referred to as photon bunching and antibunching for $g^{(2)}(0) > 1$ and $g^{(2)}(0) < 1$, respectively. When $g^{(2)}(0) = 1$ there are no correlations and the photons arrive at random.
While photon bunching is the property of photons to arrive more likely together, antibunching signatures describe the case where the probability of detecting two photons at the same time goes to zero, independent of the field intensity. The case of $g^{(2)}(0) = 0$, i.e. the absence of intensity correlations at equal time is a single photon signature~\cite{kimble1977photon}. The effect of photon antibunching received a lot of attention since it cannot be described using semi-classical methods, and therefore is a phenomena which requires the quantized electromagnetic field. The absence of joint photon detection counts is thus also used as a tool for the verification of single photon sources~\cite{scheel2009single}.

\subsection{\label{sec:intensity_correlation}Intensity correlations in HHG}

Having established the notion of classifying classical from non-classical signatures in the second order correlation function, we can proceed by looking at the intensity correlation of the harmonics. 
Using the second order correlation function from Eq.~\eqref{eq:correlation_2nd_order_general} we can write 
\begin{align}
\label{eq:gtwo_harmonic_q}
    \Gtwo (t, t+\tau) & = \hbar^4 g^4 q^2 \expval{a_q^\dagger(t) a_q^\dagger (t+\tau) a_q(t+\tau) a_q(t)} \nonumber \\
    & \equiv \hbar^4 g^4 q^2 \expval{\mathcal{N}(t,\tau)}.
\end{align}

Like in the calculations for the first order correlation, we only consider the contribution of the scattered field in Eq.~\eqref{eq:scattered_field} due to the initial vacuum state of the harmonics. For the two-time average of the normal ordered field operators, we find
\begin{align}
\label{eq:expectation_value_dipoles}
    \expval{\mathcal{N}(t,\tau)} &= g^4 q^2 \int_{t_0}^{t} dt_1 \int_{t_0}^{t+\tau} dt_2 \int_{t_0}^{t+\tau} dt_3 \int_{t_0}^{t} dt_4 \nonumber \\
    & \quad\times e^{- i \omega_q (t_1+t_2-t_3 -t_4)} \mathcal{D}(t_1,t_2,t_3,t_4),
\end{align}
where we have defined the dipole correlation term 
\begin{align}
\label{eq:dipole_correlation_2nd_order}
    \mathcal{D}(t_1,t_2,t_3,t_4) & \equiv \sum_{i,j,k,l=1}^N \Tr \left[ (\pol_q \cdot \vb{d}_i (t_1)) (\pol_q \cdot \vb{d}_j (t_2)) \right. \nonumber \\
    & \left. \quad \times (\pol_q \cdot \vb{d}_k (t_3)) (\pol_q \cdot \vb{d}_l (t_4)) \rho_S(t_0) \right],
\end{align}
where $\rho_S(t_0)$ is the initial state of the system. 
If we again only consider the dipole along the laser polarization direction, $\pol_q \cdot \vb{d}_i(t) \equiv d_i(t)$, within the $N=1$ single atom picture, we now have
\begin{align}
\label{eq:dipole_4th_order}
     \mathcal{D}(\boldsymbol{t} ) & = \bra{\text{g}} d(t_1) d(t_2) d(t_3) d(t_4) \ket{\text{g}},
\end{align}
where we have used $\rho_S(t_0) = \dyad{g}$, and introduced the tuple over the different time arguments $\boldsymbol{t} \equiv (t_1, t_2, t_3,t_4)$.
In analogy to the decomposition of the 2nd order dipole moment in Eq. \eqref{eq:dipole_single_atom}, we can write 
\begin{align}
    \mathcal{D}(\boldsymbol{t}) = & \, \expval{d(t_1) d(t_2)} \expval{d(t_3) d(t_4)} \\
    & + \expval{\Delta [d(t_1) d(t_2)] \Delta [d(t_3) d(t_4)]}, \nonumber
\end{align}
where we have defined 
\begin{align}
    \Delta [d(t_1) d(t_2)] \equiv d(t_1) d(t_2) - \expval{d(t_1) d(t_2)}.
\end{align}

We can see that the 4th order dipole moment is composed out of the product of two dipole correlation terms, and a term including the fluctuations of the dipole correlations $\Delta [d(t_1) d(t_2)]$.
Now, equipped with expressions for the second order correlation function, we can look at the normalized correlation function of a given harmonic 
\begin{align}
\label{eq:g2_definition_field_operators}
    g^{(2)}(\tau) & = \frac{\expval*{:I_q(t) I_q(t+\tau):}}{\expval*{I_q(t)} \expval*{I_q(t+\tau)}},
\end{align}
for which we normalize the second order correlations to the intensity average at the two different times. 
This can in fact be written in terms of the first and second order correlation function
\begin{align}
    g^{(2)}(\tau) = \frac{G^{(2)}(t,t+\tau)}{G^{(1)}(t+\tau) G^{(1)}(t)},
\end{align}
where we can use the results from Sec. \ref{sec:1st_order_incoherent} for the denominator 
\begin{align}
    G^{(1)} (t) = G^{(1)}_{coh}(t) + G^{(1)}_{inc}(t). 
\end{align}

\begin{figure*}
    \centering
	\includegraphics[width = \textwidth]{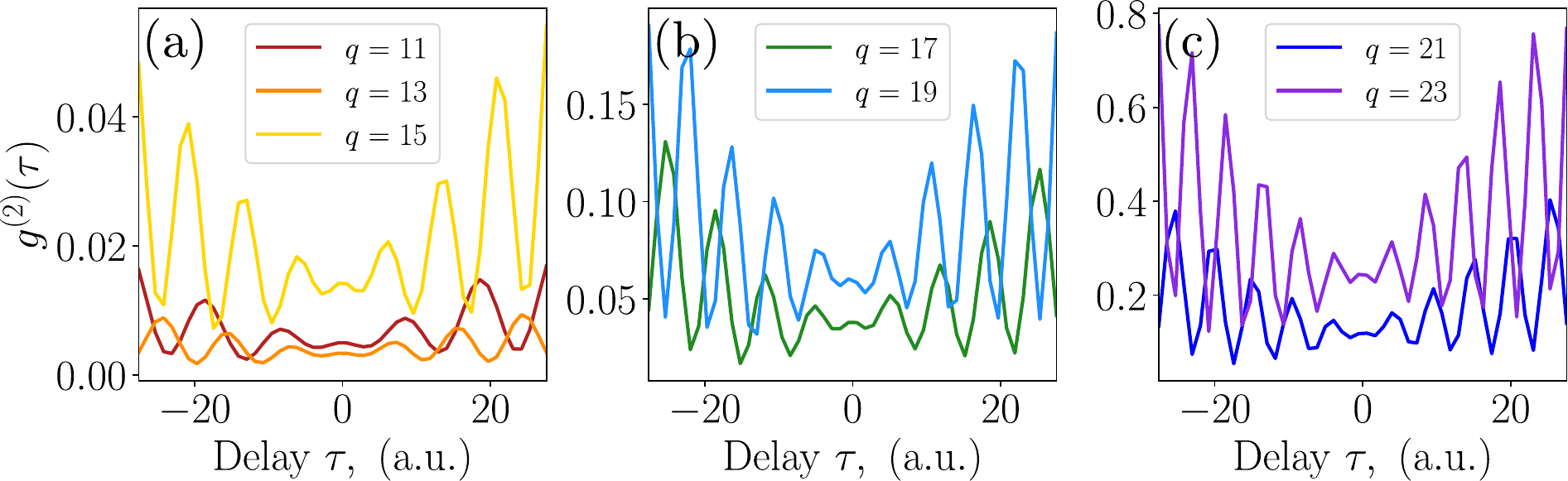}
	\caption{We show the normalized second order correlation function $g^{(2)}(\tau)$ for a few odd harmonics in the plateau (a) $q \in \{11,\, 13,\, 15 \}$  and (b) $q \in \{17, \, 19\}$, while in (c) $q \in \{21, \, 23 \}$  near the cut-off. The time-delay $\tau$ is taken to be after one complete cycle of the driving laser field $T = 2 \pi / \omega$. Here, we have used the same  driving field conditions to those in Fig.~\ref{fig:scattred_field}.}
      \label{fig:g2_function}
\end{figure*}

With the expression of the 4th order dipole moment in Eq.~\eqref{eq:dipole_4th_order}, for which a more detailed derivation is given in the Supplementary Material (SM), we can now compute the $g^{(2)}$-function for varying time-delay $\tau$. We solve the second order correlation function $G^{(2)}(t,t+\tau)$ numerically, such that all orders of the dipole moment correlations are taken into account. 
The normalized intensity correlation $g^{(2)}(\tau)$ for the harmonic modes emitted from one cycle of the driving field $t = T = 2 \pi / \omega$, are shown in Fig.~\ref{fig:g2_function} (a)-(c) for different harmonic orders. 
The first important and interesting observation is that we find photon anti-bunching across all harmonic orders, with values close to perfect single photon emission for vanishing time-delay. For instance, we have $g^{(2)}_{11}(0) \approx 5\times 10^{-3}$, and $g^{(2)}_{13}(0) \approx 3\times10^{-3}$, for the harmonics 11 and 13, respectively.  
This observation of anti-bunching signatures can also be understood from the intuitive 3-step model of the trajectory approach to HHG~\cite{corkum1993plasma, lewenstein1994theory}. Once an electron is ionized and accelerated in the continuum, it emits the harmonic photon upon recombination with the ion. Then, in order to emit another photon, the electron needs to undergo the whole 3-steps of ionization, propagation and recombination for the next emission event. Since this process is not instantaneous, the electron does not emit two photons at the same time, leading to anti-bunching signatures.  
It implies that there are hardly ever two photons emitted at the same time, independent of the driving intensity.
Furthermore, we can see a decrease in the anti-correlation measure on average (increase of the averaged $g^{(2)}$ function over time) for increasing time-delay. This indicates the expected property that for long time delays the photon correlations decrease, and will eventually become uncorrelated ($g^{(2)} = 1$) in the long time limit. 
Comparing with the classical properties of the second order correlation function, given in Eq.~\eqref{eq:g2_classical1} and Eq.~\eqref{eq:g2_classical2}, we can see that the intensity correlations obtained from HHG violate both inequalities. First, the anti-bunching signatures via $g^{(2)}(0) < 1$, and second the increase of the correlation function (in the temporal average) compared to the correlations at vanishing time-delay $g^{(2)}(0) < g^{(2)}(\tau)$. 
We further note that the second order correlation function at vanishing time-delay is closely related to the photon number distribution of the field, often compressed in the measure given by the Mandel $Q$-parameter~\cite{mandel1979sub}. The $Q$-parameter is a measure of the deviation of the variance of the photon number distribution from the variance of a Poisson process. The Poisson distribution ($Q=0$) is obtained from coherent states in which the variance of the distribution is equal to the mean. The case of $Q<0$, and $Q>0$, corresponds to sub-Poissonian and super-Poissonian distributions, respectively. Here, only the sub-Poissonian statistics corresponds to non-classical signatures in which the variance of the photon number distribution is smaller than its mean (and vice versa for super-Poissonian distributions). However, we emphasize that photon anti-bunching and sub-Poissonian statistics share common signatures, but are ultimately different and should not be confused~\cite{PhysRevA.41.475}. 

Finally, we want to stress that this is the first observation of non-classical photon correlations in the process of HHG. While previous work has shown photon bunching~\cite{lemieux2024photon}, which can be explained by classical theory, we show the first anti-bunching signatures in HHG without a classical counterpart.

\subsection{\label{sec:manyN_2nd_order}Many atom picture}

So far, we have studied the single atom scenario, and therefore only a single emitter will contribute to the emission of harmonics and intensity correlation function. Akin to the first-order correlation function, we are now interested in the many atom case. In particular because this is the regime of current HHG experiments, and consequently shows if the non-classical signatures will survive realistic conditions. In the following we will consider the case of $N$ indistinguishable emitters.

We shall first look at the higher order dipole moment expectation value in Eq. \eqref{eq:dipole_correlation_2nd_order} for the $N$ atom case along the laser polarization direction
\begin{align}
\label{eq:2nd_order_dipole_correlation_2terms}
    \sum_{i,j,k,l=1}^N &  \bra{\vb{\bar g}} d_i(t_1) d_j(t_2) d_k(t_3) d_l(t_4) \ket{\vb{\bar g}} \nonumber \\
    & = \sum_{i=1}^N \expval{d_i(t_1) d_i(t_2) d_i(t_3) d_i(t_4)} \\
    & + \sum_{i \neq \{j,k,l\} } \sum_{j,k,l,=1}^N \expval{d_i(t_1) d_j(t_2) d_k(t_3) d_l(t_4)}, \nonumber
\end{align}
where $i \neq \{ j,k,l \}$ indicates that $i$ is not equal to $j,k,l$ at the same time.
The first term is the contribution from each of the $N$ emitters individually, and the second term takes into account the correlations between different emitters. We again observe that the first term scales as $\mathcal{O}(N)$, while the second term scales as $\mathcal{O}(N^4)$. 
Now, we consider the same assumptions of Sec. \ref{sec:many_N} which includes: (i) an uncorrelated initial state of the emitters such that the dipole moment between different emitters factorize, and (ii) each emitter experiences the same driving field such that the subscript can be dropped due to the indistinguishability of the atoms. 
We thus have for the first term in Eq. \eqref{eq:2nd_order_dipole_correlation_2terms}
\begin{align}
    \sum_{i=1}^N \expval{d_i(t_1) d_i(t_2) d_i(t_3) d_i(t_4)} = N \expval{d(t_1) d(t_2) d(t_3) d(t_4)},
\end{align}
which is similar to the term evaluated in the single atom case with a linear scaling for the number of atoms $N$. 
For the second term, including the correlations between the emitters, we have 
\begin{align}
\label{eq:dipole_series_expansion}
    & \sum_{i \neq \{j,k,l\} }  \sum_{j,k,l,=1}^N \expval{d_i(t_1) d_j(t_2) d_k(t_3) d_l(t_4)} \\
    & = \frac{N!}{(N-4)!} \expval{d(t_1)} \expval{d(t_2)} \expval{d(t_3)} \expval{d(t_4)} + \mathcal{O}(N^3), \nonumber
\end{align}
where we have neglected terms of order $\mathcal{O}(N^3)$ since HHG experiments are typically performed in the $N\gg 1$ regime such that the major contribution comes from the coherent part. The coherent contribution, in which all dipole moment expectation values are evaluated independently, scales as $\mathcal{O}(N^4)$. In the SM we show the next leading terms in the expansion with respect to $N$. 
Interestingly, the coherent contribution is even more dominant in the higher order correlation functions when considering the many atom regime. This suggests that it is hard to detect non-classical signatures in the second order correlation function in the typical experimental regimes. 
However, this again holds for the uncorrelated state of the $N$ emitters, and interesting deviations from this are expected to occur when driving HHG in correlated materials \cite{pizzi2023light}, or solid state materials in which a violation of the Cauchy-Schwarz inequality was observed~\cite{theidel2024evidence, van2025errors}.

\begin{figure*}
    \centering
	\includegraphics[width = \textwidth]{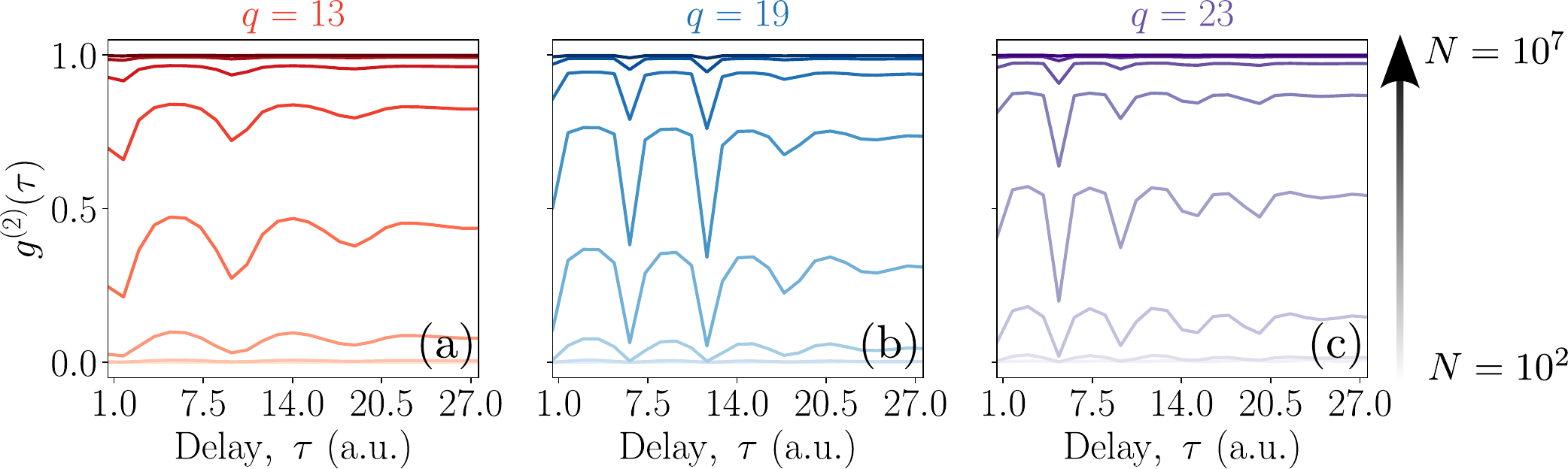}
	\caption{We show the normalized intensity correlation $g^{(2)}(\tau)$ for different harmonic orders $q= 13$ in (a), $q=19$ in (b) and $q=23$ in (c) for varying number of emitters $N$. The atom number increases for increasing color brightness including $\log_{10}N \in \{2.0, 2.7,3.4,4.1,4.8,5.5,6.7,7.0 \} $. The same field conditions as those in Fig.~\ref{fig:scattred_field} are considered here.}
      \label{fig:dipole_higher_order}
\end{figure*}

In order to discuss the influence of the many atom emission on the intensity correlation, we show in Fig.~\ref{fig:dipole_higher_order} the $g^{(2)}(\tau)$ function for different values for the number of atoms $\log_{10}N \in \{1.0,2.7,3.4,4.1,4.8,5.5,6.7,7.0\}$. As expected, the second order correlation function increases for increasing number of uncorrelated emitters until it approaches unity for very large $N$. This is attributed to the different scalings of the coherent and incoherent contributions, where the former has the $N^4$ scaling whereas the latter scales at most with $\mathcal{O}(N^3)$. The limit of $g^{(2)}(\tau) = 1$ is the well known signature of classical coherent radiation due to oscillating charge currents, here given by the dipole moment expectation value in the coherent contribution.

\section{Discussion}

In this work we have introduced the notion of optical coherence in the process of HHG, and formulated a quantum optical theory for field correlation functions. This was achieved by developing a Heisenberg picture approach for the field operators $a_q(t)$ of the harmonic modes. The Heisenberg picture allowed for a convenient way to obtain two-time averages of observables, such as field correlation functions. Since the computation of the field correlation functions did not involve specific assumptions about the HHG medium, the framework introduced in this work is applicable to all intense laser driven systems, and all driving laser configurations including structured light.

\subsection{\label{sec:conclusion}Conclusions}

With the general theory established in this work, we were able to compute field correlation functions for the process of HHG. In particular, we have seen how the HHG spectrum is constituted, and observed photon anti-correlations in HHG. These were obtained from the first and second order correlation function, respectively. For both cases we have further studied the scenario of single, few and many emitters.

In more detail, we have observed that the deviation from the classical HHG spectrum is given by the incoherent contribution originating from dipole moment fluctuations, while the coherent contribution reproduces the semi-classical result. We have further shown that these incoherent contributions are difficult to observe in conventional HHG experiments where many atoms $N$ contribute to the process. This is due to the scaling problem of the incoherent contribution compared to the coherent part, which grow as $\mathcal{O}(N)$ and $\mathcal{O}(N^2)$, respectively.
The deviation from the classical result due to the dipole correlations is in agreement with the results reported in Ref.~\cite{stammer2024entanglement}, revealing entanglement and squeezing in the process of HHG if such contributions are taken into account.  
These insights allow to take a different look at the approximations involved in previous quantum optical approaches to HHG. A variety of theoretical work has considered the case in which coherent input states are mapped to coherent output states~\cite{lewenstein2021generation, rivera2022strong, gorlach2023high}, which are related, among other approximations, to neglecting dipole moment correlations as emphasized in Sec.~\ref{sec:comparison_schrödinger} or Ref.~\cite{lange2024hierarchy}.
We have seen that in order to get the classical result, aka product coherent states in the output field, we have to assume (i) uncorrelated emitters (as shown in Sec.~\ref{sec:many_N}), and (ii) vanishing dipole moment fluctuations from a single emitter at different times~\cite{stammer2024entanglement}. 
Therefore, including dipole correlations is necessary for observing quantum signatures in the harmonic radiation, even in the single atom case.
However, the presence of many atoms leads to a coherent (classical) contribution which predominates the incoherent (quantum) contribution. This implies, that even in the presence of non-classical signatures, typical experiments are probably unable to reveal them. 

We have then studied the second order correlation function, which is related to the intensity correlation measurements in a HBT configuration~\cite{brown1956correlation}. In the $g^{(2)}(\tau)$ correlation measure, we found clear signatures of non-classicality in the photon correlations by means of anti-correlated photon emission. With this, we have reported the first indication of photon anti-bunching in the HHG emission process over a wide range of harmonic modes. This is particularly interesting because there is yet only a limited number of reported genuine quantum signatures in the HHG process, compared to many signatures explainable with classical theory (see Ref.~\cite{cruz2024quantum} for an overview on classical and quantum signatures in HHG). Reporting on photon anti-bunching paves the way to study the non-classical emission characteristics of HHG with the potential to be a versatile single photon source covering a wide range of frequencies.

\subsection{\label{sec:outlook}Outlook}

This work has introduced the quantum theory of correlation functions and optical coherence in HHG. This allows for extensive further exploration of the interplay between strong laser fields with matter from the perspective of revealing non-classical signatures in the radiation properties. Since the method in this work is complementary to the previously considered Schrödinger picture, it allows to uncover the underlying quantum signatures from a different viewpoint. Instead of focusing, and solving for the quantum state of the field, we emphasize on the observer perspective of measurable quantities, such as two-time averages and correlation functions of the field. 
This paves the way for further studies of the rapidly emerging field at the intersection of quantum optics and strong field driven dynamics and attosecond science~\cite{cruz2024quantum}. Possible future directions are outlined in the following together with an emphasis on the respective challenges.

\subsubsection{The role of the incoherent contribution}

In Sec.~\ref{sec:1st_order_incoherent}, we have seen that the incoherent contribution to the first order correlation function is the origin of non-classical signatures in the HHG spectrum (or the quantum state as shown in Ref.~\cite{stammer2024entanglement}). Since the incoherent contribution is due to the dipole moment fluctuations, we expect that analyzing such dipole correlations will provide further signatures of non-classicality in HHG. This includes, but is not limited to, resonance effects due to the presence of cavities~\cite{yi2024generation} or initial coherence in the atomic state~\cite{gauthey1995role, rivera2024squeezed}. Going beyond atomic systems has already been proven to leave imprints in the harmonic properties when considering correlated materials~\cite{pizzi2023light, lange2024electron}. Although the imprinted signatures in the field presented in Ref.~\cite{pizzi2023light} are classical, we expect to find additional quantum signatures in the field correlation functions. 
However, the challenge for experimentally witnessing these signatures remains due to the different scaling of the coherent and incoherent contribution with $N^2$ and $N$, respectively. 
From the perspective of witnessing quantum signatures it would be of great interest to experimentally separate the coherent from the incoherent contribution to the field, which respectively result in the classical and non-classical field characteristics in the correlation functions.

From a fundamental perspective, there is a mainly overlooked aspect about the emission of high order harmonic photons. It is a general consensus that, in the semi-classical picture, the harmonic emission originates from a dipole oscillator, which is backed up by the quantum optical description introduced in Ref.~\cite{lewenstein2021generation}. In Sec.~\ref{sec:1st_order} we have provided further evidence for this picture if only the coherent dipole contribution is considered. 
However, there is one piece missing in the picture. Since the initial states of the harmonic modes are in the vacuum, the generation of photons via dipole transitions can only occur from spontaneous emission, and there is no stimulated emission due to the absence of photons in this mode. Since spontaneous emission is incoherent, this is in contrast to the aforementioned coherent emission. It seems that the incoherent contribution to the scattered field can help so solve this apparent discrepancy, and answer the question whether harmonic radiation originates from dipole currents of dipole transitions.

\subsubsection{Study the correlation and coherence properties}

With the methods provided in this work, we can now study the coherence properties of the high harmonic radiation from a quantum optical perspective in more detail. 
This includes the detailed study of the precise measurement conditions and the corresponding spectra. Important aspects on that matter are the nature of averaging at the detector~\cite{van2025errors}, the precise properties of the spectral analyzers in the measurements of the spectrum~\cite{eberly1977time}, and all the subtleties when considering time-dependent and transient phenomena~\cite{eberly1980time, lu1986time, brenner1982time, eberly1977time}. These will be particularly relevant in attosecond pump-probe schemes and time-resolved analysis, taking into account the transient nature of the time-dependent processes in attosecond science. In addition, it allows to take into account the experimental settings for studying the influence of telegraph noise, such as random-phase or frequency noise from the laser fluctuations, on the spectrum~\cite{wodkiewicz1984noise, eberly1984noise}.

Furthermore, recent experiments have shown interesting signatures in the intensity cross-correlation between two different harmonics in laser driven semiconductors~\cite{theidel2024evidence, van2025errors}. These correlations show a violation of the Cauchy-Schwarz inequality and highlight the presence of entanglement between these two field modes. We expect that, within the approach established in this work, further insights into the cross correlations can be obtained. 
However, we note that all the details concerning photon detection and photon counting for ultrashort light pulses remain largely unexplored, and averages of field observables need to be obtained with care~\cite{van2025errors}.

Finally, beyond the first order coherence property of the harmonics analyzed in Sec.~\ref{sec:coherence_1st_order}, higher order coherence of the field can be studied. For instance, if second order coherence is given by $\abs{g^{(2)}(\tau)} = 1$. However, for the second order correlation function there is the imminent challenge in the ability to detect the non-classical nature by means of the photon anti-bunching signatures. The challenge in measuring a non-classical $g^{(2)}$ function lies in the presence of many atoms participating in the process of light scattering in HHG. In the example of resonance fluorescence, the observation of anti-bunching was due to the photon emission by just a single atom on average~\cite{kimble1977photon}, and the atom can only emit a single photon at a time. However, for many emitters the situation changes dramatically due to the probabilistic nature of the emission process, and each atom can emit at a different time, leading to a vanishing anti-bunching signature. The same holds for the process of HHG, and it is either a matter of isolating the photon emission from a single atom or to consider material systems in which the underlying emission dynamics is different.

\subsubsection{Generalization to arbitrary driving fields}

Recent studies have focused on inducing non-classical signatures in the harmonic radiation by using non-classical driving fields, such as squeezed states~\cite{gorlach2023high, tzur2024generation, rasputnyi2024high, rivera2024non} or photon number states~\cite{gorlach2023high, stammer2024absence}. It was shown that a squeezed driving field can induce squeezing in the harmonic field modes~\cite{tzur2024generation}, or that squeezed driving fields allow for harmonic emission in classical forbidden regimes~\cite{rivera2024non}.
In all of the aforementioned theoretical approaches, the approximations outlined in Sec.~\ref{sec:comparison_schrödinger} have been applied, by means of assuming that coherent input states are mapped to coherent output states when decomposing the arbitrary driving field in the coherent state basis. 
In contrast, the approach presented in this work is independent of such approximations and can evaluate the field expectation values for any initial driving field. However, special care needs to be taken when considering the classically driven electron dynamics, which poses limitations on the semi-classical picture when driven by non-classical light~\cite{stammer2024limitations}.

\subsubsection{Correlated materials}

The analysis in Sec.~\ref{sec:1} and \ref{sec:coherence_2nd_order} studied the first and second order correlation function in HHG, respectively, and considered the case of atomic gas targets. In the respective subsection \ref{sec:many_N} and \ref{sec:manyN_2nd_order} the many atom case was discussed, where we have assumed uncorrelated emitters (which is usually the case for atomic gas targets). Since we have seen that the interesting departures from the classical results appear due to dipole moment correlations, we emphasize on the importance of studying correlated materials. In particular, solid state systems featuring strong correlations have been shown to leave its signatures in the quantum optical state of HHG~\cite{pizzi2023light, lange2024electron}.
We therefore expect that field correlation functions are sensitive to the correlations or non-trivial topology of the matter system driven by intense light fields.

\begin{acknowledgments}

P.S. would like to express his gratitude to Alexander Carmele for the introduction to quantum optical methods and the perspetive of the observer.
P.S. acknowledges funding from the European Union’s Horizon 2020 research and innovation programe under the Marie Skłodowska-Curie grant agreement No 847517.~ICFO group acknowledges support from: Ministerio de Ciencia y Innovation Agencia Estatal de Investigaciones (R$\&$D project CEX2019-000910-S, AEI/10.13039/501100011033, Plan National FIDEUA PID2019-106901GB-I00, FPI), Fundació Privada Cellex, Fundació Mir-Puig, and from Generalitat de Catalunya (AGAUR Grant No. 2017 SGR 1341, CERCA program), and MICIIN with funding from European Union NextGenerationEU(PRTR-C17.I1) and by Generalitat de Catalunya and EU Horizon 2020 FET-OPEN OPTOlogic (Grant No 899794) and ERC AdG NOQIA.

\end{acknowledgments}

\bibliography{literatur}{}


\newpage
\appendix

\begin{center}
    \textbf{APPENDIX}
\end{center}

\section{\label{app:details_analytic}Details on the Heisenberg equations of motion}

\subsection{\label{app:heisenberg_field}Derivation of the Heisenberg equation of motion for the field operator}

The equation of motion of the annihilation operator $a_{q, \lambda}(t)$ in \eqref{eq:aq_heisenberg_eom} can be obtained by using the Heisenberg equation 
\begin{align}
\label{eq:app:heisenberg_EOM}
    \dv{t} a_{q, \lambda}(t) & = \frac{i}{\hbar} \left[ H(t), a_{q, \lambda}(t) \right] \\
    & = \frac{i}{\hbar} U^\dagger(t) \left[ H, a_{q, \lambda} \right] U(t) , 
\end{align}
where the Hamiltonian $H$ is given in~\eqref{eq:hamiltonian_total}. We now need to evaluate the commutator
\begin{align}
\label{eq:commutation_relation}
    \left[ H, a_{q, \lambda} \right] = \left[ H_F + H_I, a_{q, \lambda} \right],
\end{align}
with the field and interaction Hamiltonian given by 
\begin{align}
    H_F & = \sum_{\lambda=1}^2 \sum_{q=1}^{q_c} \hbar \omega_{q, \lambda} a_{q, \lambda}^\dagger a_{q, \lambda},\\
    H_I & = - \sum_{i=1}^N \vb{d}_i \cdot \vb{E}_Q,
\end{align}
where we have the electric field operator
\begin{align}
    \vb{E}_Q = - i \tilde g \sum_{q, \lambda} \sqrt{q} \epsilon_{q, \lambda} \left( a_{q, \lambda}^\dagger - a_{q, \lambda} \right).
\end{align}

Using the commutation relation for the bosonic field operators 
\begin{align}
    \comm{a_{q, \lambda}}{a_{p, \lambda^\prime}^\dagger} = \delta_{\lambda, \lambda^\prime} \delta_{qp},
\end{align}
we find that the commutator in \eqref{eq:commutation_relation} is given by 
\begin{align}
    \comm{H_F + H_I}{a_{q, \lambda}} = - \hbar \omega_{q, \lambda} a_{q, \lambda} - i \tilde g \sum_{i=1}^{N} \sqrt{q} \epsilon_{q, \lambda} \cdot \vb{d}_i.
\end{align}

Inserting this into the Heisenberg equation of motion in Eq.~\eqref{eq:app:heisenberg_EOM}, we obtain 
\begin{align}
    \dv{t} a_{q, \lambda}(t) & = - i \omega_{q, \lambda} U^\dagger (t) a_{q, \lambda} U(t) \\
    & \quad + g \sqrt{q} \sum_{i=1}^N \epsilon_{q, \lambda} \cdot U^\dagger (t) \vb{d}_i U(t) \nonumber \\
    & = - i \omega_{q, \lambda} a_{q, \lambda}(t) + g \sqrt{q} \sum_{i=1}^N \epsilon_{q, \lambda} \cdot \vb{d}_i(t),
\end{align}
where $\vb{d}_i (t) = U^\dagger (t) \vb{d}_i U(t)$ is the dipole moment in the Heisenberg picture, and $g \equiv \tilde g / \hbar = \sqrt{\omega/ (2 V \hbar \epsilon_0)}$.
The solution to this differential equation is given by 
\begin{align}
    a_{q, \lambda}(t) = &  a_{q, \lambda}(0) e^{- i \omega_{q, \lambda} t } \\
    & \quad + g \sqrt{q} \int_0^t dt' e^{- i \omega_{q, \lambda} (t-t')} \sum_{i=1}^N \epsilon_{q, \lambda} \cdot \vb{d}_i(t'), \nonumber
\end{align}
which coincides with Eq.~\eqref{eq:aq_heisenberg_solution} of the main text. We note that the field operator at the initial time in the Heisenberg picture corresponds to the operator in the Schrödinger picture $a_{q, \lambda}(0) \equiv a_{q, \lambda}$.

\subsection{\label{app:classical_interaction}Heisenberg picture for the dipole moment with the semi-classical interaction}

For the dynamics of the time-dependent dipole moment $\vb{d}_i(t)$ in Eq.~\eqref{eq:aq_heisenberg_solution} we consider the corresponding Heisenberg EOM 
\begin{align}
    \dv{t} \vb{d}_i(t) = \frac{i}{\hbar} U^\dagger(t) \left[ H, \vb{d}_i \right] U(t),  
\end{align}
where we again have the Hamiltonian $H = H_F + H_S + H_I$ from Eq.~\eqref{eq:hamiltonian_total}. In order to evaluate the commutator with the Hamiltonian we first introduce a suitable reference frame via unitary transformation into the co-rotating frame of $H_F$, and of the initial coherent state $\ket{\alpha}$ of the driving laser field to the vacuum by using the unitary displacement operator $D^\dagger(\alpha) \ket{\alpha} = \ket{0}$. Thereby, we obtain an additional semi-classical interaction Hamiltonian (compare with Eq.~\eqref{eq:interaction_semiclassical} of the main text) 
\begin{align}
    H_{I,cl}(t) = - \sum_{i=1}^N \vb{d}_i \cdot \vb{E}_{cl}(t),
\end{align}
where the classical electric field is given by $\vb{E}_{cl}(t) = \Tr[\vb{E}_Q(t) \rho(0)]$, with the total Hamiltonian in the new frame 
\begin{align}
    H = H_S + H_I + H_{I,cl}.
\end{align}

Now, we are in a position to perform approximations based on the situation under consideration. Since we are interested in the case of intense driving laser fields, the interaction with the classical electric field is much larger than the interaction with the quantized field. Therefore, we only consider the semi-classical interaction $H_{I,cl}(t)$, and omit the contribution from the backaction of the quantized interaction on the electron dynamics. Thus, the Heisenberg EOM reduce to 
\begin{align}
    \dv{t} \vb{d}_i(t) = \frac{i}{\hbar} U^\dagger(t) \left[ H_S + H_{I,cl}(t), \vb{d}_i \right] U(t),  
\end{align}
which is the dynamics governed by the well-known semi-classical Hamiltonian $H_{sc}(t) = H_S + H_{I,cl}(t)$. Therefore, instead of explicitly solving the Heisenberg EOM for the dipole moment, we consider the generic solution under the semi-classical Hamiltonian 
\begin{align}
    \vb{d}_i(t) = U^\dagger_{sc}(t) \vb{d}_i(0) U_{sc}(t),
\end{align}
where we have defined the semi-classical propagator 
\begin{align}
    U_{sc}(t) = \mathcal{T} \exp[- \frac{i}{\hbar} \int_{t_0}^t dt^\prime H_{sc}(t^\prime) ].
\end{align}

This corresponds to the solution Eq.~\eqref{eq:dipole_semi-classical} of the main text. 

\newpage
\onecolumngrid
\setcounter{section}{0}
\include{Supplementary_for_main}

\end{document}

%% file: Supplementary_for_main.tex
\begin{center}
        \textbf{SUPPLEMENTARY MATERIAL}
\end{center}


\section{Computing the scattered field and the coherent contribution to the HHG spectrum}

The coherent contribution to the HHG spectrum both require to compute the Fourier transform of the time-dependent dipole moment (we will assume linear polarization in the following), i.e.
\begin{equation}
	\langle d(\omega) \rangle
		= \int^t_{t_0}
				\dd t_2\ e^{-i\omega t_2} \langle \hat{d}(t_2)\rangle,
\end{equation}
for which, in atomic units ($\hbar =1, m_{\mathsf{e}}=1, \abs{\mathsf{e}}=1$), the time-dependent dipole moment reads as~\cite{lewenstein1994theory,olga_simpleman};
\begin{equation}\label{Eq:dipole}
	\langle \hat{d}(t_2)\rangle
		= i \int \dd p 
				\int^{t_2}_{0} \dd t_2
					\langle \text{g}\vert \hat{d}\vert p+A_{\text{cl}}(t_2)\rangle
					e^{-iS(p,t_2,t_1)}E_{\text{cl}}(t_1)
					\langle p+A(t_1)\vert \hat{d}\vert \text{g}\rangle
					+ \text{c.c.},
\end{equation}
where $S(p,t_2,t_1)$ denotes the semi-classical action
\begin{equation}
	S(p,t_2,t_1)
		= \dfrac12 
			\int^{t_2}_{t_1}
				\dd \tau
					 \big[
					 	p + A_{\text{cl}}(\tau)
					 \big]^2
					  + I_p(t_2-t_1).
\end{equation}

In the numerical analysis, we used  the following formula for the matrix elements of the dipole operator taken for hydrogenic 1s-type of orbitals~\cite{podolsky_momentum_1929};
\begin{equation}\label{Eq:dipole:mel}
	\vb{d}(\vb{p})
		= \langle \vb{p}\vert \hat{\vb{d}}\vert \text{g}\rangle
		= \dfrac{8i}{\pi}
			\dfrac{\sqrt{2\kappa^5} \vb{p}}{(\vb{p}^2 + \kappa^2)^2},
\end{equation}
where $I_p = 2\kappa^2$.
The steps in the numerical analysis are as follows:
\begin{itemize}
	\item First, a grid of $N_{\text{els}} = 2000$ elements of momentum values $p$, was defined within the range $p \in [-3,3]$. 
	\item For each value of $p$, the time integral in Eq.~\eqref{Eq:dipole} was computed using the \texttt{quad} function of the \texttt{Scipy} package. The electric field was defined as $E_{\text{cl}}(t) = E_0 \cos(\omega_L t)$, where $E_0 = 0.053$ a.u. and $\omega_L = 0.057$ a.u., corresponding to an intensity of $I = 10^{14}$ W/cm$^2$ and a wavelength $\lambda = 800$ nm, respectively. The algorithm implemented by the \texttt{quad} used $1000$ iterations, which was enough to ensure convergence of the integrals. The time integration was performed over the range $t_2 \in [0,16\pi/\omega_L]$, corresponding to a total of 8 optical cycles. The time grid was set with $2\lceil 16\pi/\omega_L \rceil$ elements.
	\item After solving the integral for each $p$, denoted as $d(p,t)$, the integral over momentum was approximated using the trapezoidal method
	\begin{equation}
		\langle \hat{d}(t_2)\rangle
			= \int \dd p \ d(p,t)
			\approx \sum_{i=0}^{2000}d(p_i,t)\Delta p,
	\end{equation}
	where $\Delta p = \abs{p_{i+1}-p_i}$ was chosen to be a constant.
\end{itemize}

\begin{figure}
	\centering
	\includegraphics[width = 1\textwidth]{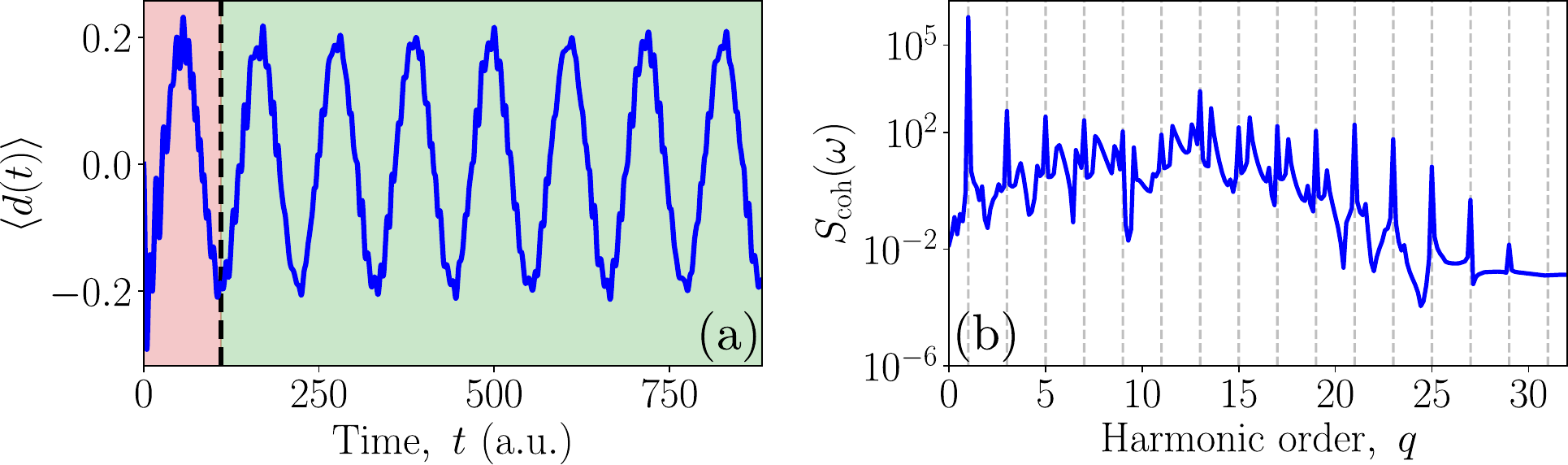}
	\caption{In (a), the dipole computed with the steps mentioned in the text. The red region corresponds to the warm up part of the integral, while the green region is the one we kept for computing the HHG spectrum shown in (b). Although initially we considered a total of 8 optical cycles, the HHG spectrum uses a total of 7 optical cycles.}
	\label{Fig:dipole}
\end{figure}

The results of this analysis are shown in Fig.~\ref{Fig:dipole}~(a), where two distinct regions can be observed: a red-shaded region corresponding to a \emph{warm-up} period where the integral does not exhibit the periodic structure observed later. This region was neglected in subsequent calculations. For later calculations, only the results starting from one optical cycle after the warm-up period were considered, effectively working with a field composed of 7 optical cycles

The dipole shown in Fig.~\ref{Fig:dipole}~(a) was subsequently used to compute other physical observables, such as the HHG spectra. To achieve this, the computed data---consisting of $2\lceil 16\pi/\omega_L \rceil$ points---was interpolated using the \texttt{interp1d} function of the \texttt{Scipy} package. This allowed for a smooth representation of the dipole moment over time. This interpolated function was used to perform a Fast Fourier Transform (FFT) within the green dashed region in Fig.\ref{Fig:dipole}~(a), using a total of $N_{\text{FFT}}=10000$ points. The resulting HHG spectrum is shown in Fig.~\ref{Fig:dipole}~(b). The spectrum exhibits peaks located at odd harmonic orders. In the plateau region, additional contributions from other frequencies are noticeable, while beyond the cutoff (approximately the 21st harmonic order), the harmonic peaks become more sharply defined.

\begin{figure}
	\centering
	\includegraphics[width = 1\textwidth]{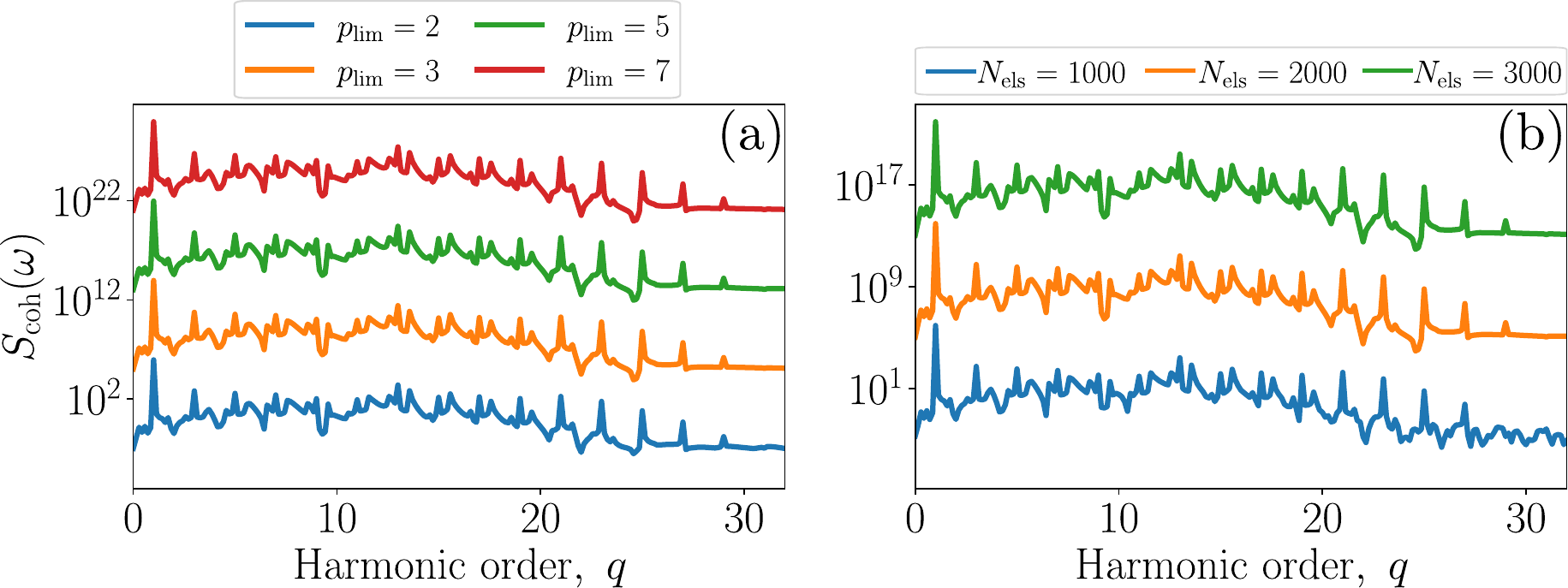}
	\caption{In (a), the HHG spectrum has been computed for $p\in [-p_{\text{lim}},p_{\text{lim}}]$, where the number of elements in the range has been set to $N_{\text{els}} = 1000$ for $p_{\text{lim}} = 2$, $N_{\text{els}} = 3000$ for $p_{\text{lim}} = 3$, $N_{\text{els}} = 5000$ for $p_{\text{lim}} = 5$ and $N_{\text{els}} = 7000$ for $p_{\text{lim}} =7$. In (b), the value of $p_{\text{lim}}$ has been fixed to 3, while the values of $N_{\text{els}}$ change in each case.}
	\label{Fig:Comparison:els}
\end{figure}

This calculation was repeated using different definitions of the $p$ interval, either by modifying the upper and lower limits or by changing the number of elements within the interval. Specifically, we selected ranges that yielded convergent results, even for shallow regions of $p$. This analysis is shown in Fig.~\ref{Fig:Comparison:els}~(a), where both the limits of the range $p\in [-p_{\text{lim}},p_{\text{lim}}]$ and the number of elements $N_{\text{els}}$ are varied. In panel (b), the upper and lower limits were fixed at $p_{\text{lim}} = \pm3$, while only the number of elements $N_{\text{els}}$ iwas varied.

\section{Computing the incoherent part of the spectrum}
The incoherent part of the spectrum is defined as
\begin{equation}\label{Eq:Inc:contrib}
	S_{\text{inc}}(\omega)
		\propto \int \dd v\
			 d_{gv}(\omega_q)d^*_{gv}(\omega_q) \delta_{\omega,\omega_q},
\end{equation}
where we have that
\begin{equation}
	d_{gv}(\omega)
		= \int^{\infty}_{-\infty} \dd t \
				\langle \text{g} \vert \hat{d}(t) \vert v \rangle e^{-i\omega_q t},
\end{equation}
where, in turn, and according to Refs.~\cite{stammer2023quantum,stammer2024entanglement};
\begin{equation}\label{Eq:dgv}
		\langle \text{g} \vert \hat{d}(t) \vert v \rangle
			=  e^{-iS(p,t,t_0)} \langle \text{g}\vert \hat{d}\vert p+A_{\text{cl}}(t)\rangle.
\end{equation}

The numerical procedure used to compute the incoherent part of the HHG spectrum is as follows:
\begin{itemize}
	\item For each value of $p\in [-p_{\text{lim}},p_{\text{lim}}]$ (using $N_{\text{els}} = 2000$ uniformly distributed points, consistent with the values used in Fig.~\ref{Fig:dipole}~(a)), we compute the $d_{gv}(t)$ function according to Eq.~\eqref{Eq:dgv}.
	\item For each $d_{vg}(t)$, we perform its FFT (prior interpolation of the function) with $N_{\text{FFT}}=10000$ points. The corresponding $d^*_{gv}(\omega)$ was obtained by performing the complex conjugate of the computed values. The incoherent contribution to the HHG spectrum was then calculated by performing the integral in Eq.~\eqref{Eq:Inc:contrib} using the trapezoidal method, similar to the approach for the coherent contribution.
\end{itemize}

Here, we chose $p_{\text{lim}} = 3$, as it was sufficient enough to find convergence of the results. However, it is worth highlighting that different values were chosen for this value in the second-order autocorrelation function analysis ($p_{\text{lim}} = 5$ and $N_{\text{els}} = 2500$), as were needed for better convergence of the integrals.

\section{Analysis of the second-order correlation functions}

The numerical methods employed to compute the second-order autocorrelation function stem as direct extension to those used for the analysis of the coherent and incoherent contributions to the spectra. The main difference stems on the particular expressions under consideration. Here, we focus on providing a more explicit version of the (unnormalized) second-order autocorrelation function. 

\subsection{Further analysis}

We start with the exact expression of Eq.~(72) in the main manuscript
\begin{align}
    \expval{\mathcal{N}(t,\tau)} &= g^4 q^2 \int_{t_0}^{t} dt_1 \int_{t_0}^{t+\tau} dt_2 \int_{t_0}^{t+\tau} dt_3 \int_{t_0}^{t} dt_4  e^{- i \omega_q (t_1+t_2-t_3 -t_4)} \expval{d(t_1) d(t_2) d(t_3) d(t_4)}.
\end{align}

First, we substitute $t_i = t_i - \tau$ for $i = 2,3$, such that 
\begin{align}
    \expval{\mathcal{N}(t,\tau) } &= g^4 q^2 \int_{t_0}^{t} dt_1 \int_{t_0}^{t} dt_2 \int_{t_0}^{t} dt_3 \int_{t_0}^{t} dt_4  e^{- i \omega_q (t_1+t_2-t_3 -t_4)} \expval{d(t_1) d(t_2+\tau) d(t_3+\tau) d(t_4)},
\end{align}
the different lower bound $t_0 - \tau$ is omitted and replaced with $t_0$ since the laser is off before the initial time when $t_0 \to \infty$. In the exponential the $\tau$ cancels due to the different sign for the $t_2$ and the $t_3$ integration. 
We do now consider the higher order dipole expectation value, and use the cyclic property of the trace (with $\rho_0 = \dyad{g}$ the initial atomic state);
\begin{align}
    &\Tr[d(t_1) d(t_2+\tau) d(t_3+\tau) d(t_4) \rho_0] = \Tr[ d(t_4) \rho_0 d(t_1) d(t_2 + \tau) d(t_3+\tau) ] \\
    & = \bra{g} d(t_4) \dyad{g} d(t_1) d(t_2+ \tau) d(t_3+\tau) \ket{g} + \int \dd v \bra{v} d(t_4) \dyad{g} d(t_1) d(t_2+ \tau) d(t_3+\tau) \ket{v} \\
    & = \bra{g} d(t_4) \dyad{g} d(t_1) \dyad{g} d(t_2+\tau) d(t_3+\tau) \ket{g} + \int \dd v \bra{g} d(t_4) \dyad{g} d(t_1) \dyad{v} d(t_2+\tau) d(t_3+\tau) \ket{g} \\
    & + \int \dd v \bra{v} d(t_4) \dyad{g} d(t_1) \dyad{g} d(t_2+\tau) d(t_3+\tau) \ket{v} + \int \dd v \int \dd v' \bra{v} d(t_4) \dyad{g} d(t_1) \dyad{v'} d(t_2+\tau) d(t_3+\tau) \ket{v}.
\end{align}

Considering the full expression from above, we find (omitting the $g^4 q^2$ pre-factor)
\begin{align}
\label{eq:g2_analysis_appendix1}
    \expval{\mathcal{N}(t,\tau)} = & \quad G^{(1)}_{coh}(t) G^{(1)}(t+\tau) \\
\label{eq:g2_analysis_appendix2}
    & + \int_{t_0}^{t} dt_1 \int_{t_0}^{t} dt_2 \int_{t_0}^{t} dt_3 \int_{t_0}^{t} dt_4  e^{- i \omega_q (t_1+t_2-t_3 -t_4)} 2 \operatorname{Re} \left[ \int \dd v \bra{g} d(t_4) \dyad{g} d(t_1) \dyad{v} d(t_2+\tau) d(t_3+\tau) \ket{g} \right] \\
\label{eq:g2_analysis_appendix3}
    & + \int_{t_0}^{t} dt_1 \int_{t_0}^{t} dt_2 \int_{t_0}^{t} dt_3 \int_{t_0}^{t} dt_4  e^{- i \omega_q (t_1+t_2-t_3 -t_4)} \int \dd v \int \dd v' \bra{v} d(t_4) \dyad{g} d(t_1) \dyad{v'} d(t_2+\tau) d(t_3+\tau) \ket{v}.
\end{align}

Note that in the first term we have for $ G^{(1)}(t+\tau) = G^{(1)}_{coh}(t+\tau) + G^{(1)}_{inc}(t+\tau)$, the coherent and incoherent contribution, respectively.
Here we are interested in the sub-cycle intensity correlations, and we therefore consider a "quasi stationary limit" in which the harmonic emission is already build up by considering $t = T = 2 \pi / \omega$ being a full cycle of the field. 
More specifically, we focus on Eqs.~\eqref{eq:g2_analysis_appendix2} and \eqref{eq:g2_analysis_appendix3} of the main text, as Eq.~\eqref{eq:g2_analysis_appendix1} depends on the first-order autocorrelation function which we already know how to compute.

\subsection{Analysis of Eq.~\eqref{eq:g2_analysis_appendix3}}
We begin this analysis by looking at Eq.~\eqref{eq:g2_analysis_appendix3}, which is the more complicated one. The lessons learned here will then be applied to compute Eq.~\eqref{eq:g2_analysis_appendix2}. From~\eqref{eq:g2_analysis_appendix3}, we are mostly interested in the momentum integrations as they are the ones we can simplify the most. More specifically, we can write
\begin{equation}
	\begin{aligned}
	f(t_1,t_2,t_3,t_4) 
		&= \int \dd v_1 \int \dd v_2
				d_{v_1g}(t_4)
				d^*_{v_2g}(t_1)
				\langle v_2 \vert d(t_2)d(t_3)\vert v_1 \rangle
		\\&
		= \int \dd v_1 \int \dd v_2
				d_{v_1g}(t_4)
				d^*_{v_2g}(t_1)
				d_{v_2,g}(t_2)
				d^*_{v_1,g}(t_3)
			+ \int \dd v_1 \int \dd v_2 \int \dd v_3
					d_{v_1g}(t_4)
					d^*_{v_2g}(t_1)
					d_{v_2,v_3}(t_2)
					d_{v_3,v_1}(t_3),
	\end{aligned}
\end{equation}
where the continuum-continuum matrix elements can be written as~\cite{stammer2023quantum};
\begin{equation}\label{Eq:cc:time}
	d_{v_2,v_3}(t_2)
		= \int \dd v \int \dd v'
				b^*_{v_2}(v,t_2) b_{v_3}(v',t_2)
				\langle v\vert r \vert v' \rangle,
\end{equation}
with $b_v(v',t)$ denoting the probability amplitude of finding an electron in the continuum with kinetic momentum $v'$, when it initially had momentum $v$. More explicitly, its expression is given by
\begin{equation}\label{Eq:b:prob:amp}
	b_v(v',t) = e^{-iS(v',t,t_0)} \delta(v'-v).
\end{equation}

Inserting Eq.~\eqref{Eq:b:prob:amp} into Eq.~\eqref{Eq:cc:time} and applying the chain rule, we arrive to
\begin{equation}\label{Eq:cc:term:expanded}
	d_{v_2,v_3}(t_2)
		= \Delta r(v_2,t_2,t_0) \delta(v_2 - v_3)
			+ i \partial_{v_2}\big[ \delta(v_2-v_3)\big],
\end{equation}
such that the whole expression explicitly reads
\begin{align}
	I &=
		\int \dd v_1\! \int \dd v_2\! \int \dd v_3\
			d_{v_1,g}(t_4) d^*_{v_2,g}(t_1)
			\Big[
				\Delta r(v_2,t_2,t_0) \delta(v_2 - v_3)
				+ i \partial_{v_2}\big[ \delta(v_2-v_3)\big]\nonumber
			\Big]
			\\&\hspace{3.3cm}\times
			\Big[
				\Delta r(v_3,t_3,t_0) \delta(v_3 - v_1)
				+ i \partial_{v_3}\big[ \delta(v_3-v_1)\big]
			\Big]
		\\&
		=\int \dd v_1\! \int \dd v_2\! \int \dd v_3\
					d_{v_1,g}(t_4) d^*_{v_2,g}(t_1)
					\Delta r(v_2,t_2,t_0)
					\Delta r(v_3,t_3,t_0)
					\delta(v_2-v_3)\delta(v_3-v_1)	\label{Eq:int1}
		\\&\quad
			+ i\int \dd v_1\! \int \dd v_2\! \int \dd v_3\
					d_{v_1,g}(t_4) d^*_{v_2,g}(t_1)
					\Delta r(v_2,t_2,t_0)
					\delta(v_2-v_3)
					\partial_{v_3}\big[\delta(v_3-v_1)\big]	\label{Eq:int2}
		\\&\quad
			+ i\int \dd v_1\! \int \dd v_2\! \int \dd v_3\
					d_{v_1,g}(t_4) d^*_{v_2,g}(t_1)
					\Delta r(v_3,t_3,t_0)
					\delta(v_3-v_1)
					\partial_{v_2}\big[\delta(v_2-v_3)\big]	\label{Eq:int3}
		\\&\quad
			-\int \dd v_1\! \int \dd v_2\! \int \dd v_3\
					d_{v_1,g}(t_4) d^*_{v_2,g}(t_1)
					\partial_{v_3}\big[\delta(v_3-v_1)\big]
					\partial_{v_2}\big[\delta(v_2-v_3)\big]	\label{Eq:int4}.
            \\& \equiv I_1 + I_2 + I_3 + I_4.
\end{align}

Let us evaluate each of the terms in Eqs.~\eqref{Eq:int1} to \eqref{Eq:int4} separately. For the first, we trivially get
\begin{equation}
	I_1
		= \int \dd v_3 \
				d_{v_3,g}(t_4)
				d^*_{v_3,g}(t_1)
				\Delta r(v_3,t_2,t_0)
				\Delta r(v_3,t_3,t_0),
\end{equation}
which in case of setting al values of $t$ to be equal, we find it is a real-valued quantity since $I_1 = I_1^*$.

For Eq.~\eqref{Eq:int2}, we first perform the integral over $v_2$, as we get a Dirac delta without any kind of derivatives in front
\begin{align}
	I_2
		= i\int \dd v_1 \int \dd v_3\
				d_{v_1,g}(t_4) d^*_{v_3,g}(t_1)
				\Delta r(v_3,t_2,t_0)
				\partial_{v_3}\big[\delta(v_3-v_1)\big],
\end{align}
and next, we follow the integration order and do the integral with respect to $v_3$ applying the integration by parts method. This results in
\begin{align}
	I_2
		= -i \int \dd v_1\
				\partial_{v_1}
				d_{v_1,g}(t_4)
				\big[
					d^*_{v_1,g}(t_1)
					\Delta r(v_1,t_2,t_0)
				\big].
\end{align}

For Eq.~\eqref{Eq:int3}, we start by performing the integral with respect to $v_1$, for similar reasons as before
\begin{equation}
	\begin{aligned}
	I_3 &= i \int \dd v_2 \int \dd v_3\
		   	d_{v_3,g}(t_4)d_{v_2,g}^*(t_1)
		   	\Delta r(v_3,t_3,t_0)
		   	\partial_{v_2}\big[\delta(v_2-v_3)\big]
		   	\\&= i \int \dd v_2\ d_{v_2,g}^*(t_1) \int \dd v_3\
		   	d_{v_3,g}(t_4)
		   	\Delta r(v_3,t_3,t_0)
		   	\partial_{v_2}\big[\delta(v_2-v_3)\big]
		   	\\& 
		 	= i\int \dd v_2\
		 		d^*_{v_2,g}(t_1)
		 		\partial_{v_2}
		 			\big[
		 				d_{v_2,g}(t_4)
		 				\Delta r(v_2,t_3,t_0)
		 			\big],
	\end{aligned}
\end{equation}
where in moving from the first to the second equality we took the derivative with respect to $v_2$ out, and then integrate over $v_3$. As we have seen, if we set all times to be equal, we observe that $I_2 = I_3^*$. Although this is not the case, later on we have to perform some integrals with respect to time, which will make the term real-valued.

Finally, for Eq.~\eqref{Eq:int4}, we integrate first over $v_1$, which is the integral for which we have a Dirac delta only. This leads to
\begin{equation}
	I_4
		= - \int \dd v_2 \int \dd v_3\
				 d^*_{v_2,g}(t_1) 
				 \partial_{v_3}
				 	\big[ d_{v_3,g}(t_4)\big]
				\partial_{v_2} \big[ \delta(v_2-v_3)\big],
\end{equation}
and now we do the integral over $v_2$, to get
\begin{equation}
	I_4
		= \int \dd v_3\
				\partial_{v_3}
				\big[d_{v_3,g}(t_4)\big]
				\partial_{v_3}
				\big[ d^*_{v_3,g}(t_1)\big].
\end{equation}

\subsection{Analysis of Eq.~\eqref{eq:g2_analysis_appendix2}}
From the analysis we have done thus far, analytical expression for Eq.~\eqref{eq:g2_analysis_appendix2} are more straightforward. From this contribution, we are only interested in the real part, although here we evaluate the complete integral without distinguishing between real or imaginary parts, and make this distinction in the numerical analysis. The momentum integral in this case is given by
\begin{equation}
	\begin{aligned}
	I &= \int \dd v_1
			 d(t_4) d^*_{v_1,g}(t_1)
			 \bra{v_1}{d}(t_2){d}(t_3)\ket{g}
	  \\&= \int \dd v_1
	  		d(t_4) d^*_{v_1,g}(t_1)
	  		d_{v_1,g}(t_2)d(t_3)
	  	+ \int \dd v_1 d(t_4) d^*_{v_1,g}(t_1)
	  		\int \dd v_2
	  			d_{v_1,v_2}(t_2)
	  			d_{v_2,g}(t_3),
	\end{aligned}
\end{equation}
where in going from the first to the second equality we have introduced the identity. In the second equality, the first term is computed numerically, while for the second one we have to do some analytics first to reduce the number of integrals. Specifically, using Eq.~\eqref{Eq:cc:term:expanded}, we can write the integral with respect to $v_2$ as follows
\begin{align}
	\int \dd v_2 d_{v_1,v_2}(t_2) d_{v_2,g}(t_3)
		&= \Delta r(v_1,t_2,t_0) d_{v_1,g}(t_3)
			+ i \int \dd v_2 d_{v_2,g}(t_3)\partial_{v_1}\big[ \delta(v_1-v_2)\big]
		\\&
		=  \Delta r(v_1,t_2,t_0) d_{v_1,g}(t_3)
		+ i \partial_{v_1}
			\bigg[
				\int \dd v_2 d_{v_2,g}(t_3) \delta(v_1-v_2)
			\bigg]
		\\&
		= \Delta r(v_1,t_2,t_0) d_{v_1,g}(t_3)
			+i \partial_{v_1} \big[d_{v_1,g}(t_3)\big],
\end{align} 
and therefore the total integral reads
\begin{equation}
	\begin{aligned}
	I &= \int \dd v_1
			d(t_4) d^*_{v_1,g}(t_1)
			d_{v_1,g}(t_2)d(t_3)
		+ \int \dd v_1
				 d(t_4) d^*_{v_1,g}(t_1)
				 \Delta r(v_1,t_2,t_0) d_{v_1,g}(t_3)
		+i \int \dd v_1
				d(t_4) d^*_{v_1,g}(t_1)
				  \partial_{v_1}\big[d_{v_1,g}(t_3)\big].
	\end{aligned}
\end{equation}

\subsection{Normalized second order correlation function}

A more useful measure of the second order correlation function (the intensity correlation function) is the normalized version of it. This has the advantage of being a dimensionless measure independent of the order of magnitude of the intensities. Therefore, we consider 
\begin{align}
    g^{(2)} (\tau) = \frac{ \expval{: I(t) I(t+\tau)  :} }{ \expval{I(t)} \expval{I(t+\tau)} } = \frac{ \expval{a_q^\dagger(t) a_q^\dagger(t+\tau) a_q(t+\tau) a_q(t)} }{ \expval*{a_q^\dagger(t) a_q(t)} \expval*{a_q^\dagger(t+\tau) a_q(t+\tau)} }.
\end{align}

The enumerator we have discussed above, and found expressions ready to be solved numerically. The denominator is simple, and known already since it is given by the first order correlation function at equal time. We therefore have 
\begin{align}
    \expval{I(t)} = \expval{a_q^\dagger (t) a_q(t)} = G^{(1)}(t,t) = \hbar^2 g^4 q^2 \left[ \abs{ \int_{t_0}^t dt_1 e^{- i \omega_q t_1} \expval{d(t_1)} }^2 + \int dv d_{gv}(\omega_q) d_{vg}^*(\omega_q)  \right],
\end{align}
where 
\begin{align}
    d_{gv}(\omega_q) = \int_{t_0}^t dt_q d_{gv}(t_1) e^{- i \omega_q t_1}.
\end{align}

\section{\label{app:many_atom} Higher order dipole moment correlations in the many atom regime}

The next leading terms in Eq.~(82) are given by 
\begin{align}
    \sum_{i \neq \{j,k,l\} } & \sum_{j,k,l,=1}^N  \expval{d_i(t_1) d_j(t_2) d_k(t_3) d_l(t_4)} =  \frac{N!}{(N-4)!} \expval{d(t_1)} \expval{d(t_2)} \expval{d(t_3)} \expval{d(t_4)} \\
    & + \frac{N!}{(N-3)!} \left[ \expval{d(t_3)} \expval{d(t_4)} \expval{d(t_1) d(t_2)}  \right.  + \expval{d(t_2)} \expval{d(t_4)} \expval{d(t_1) d(t_3)} \\
    & + \expval{d(t_2)} \expval{d(t_3)} \expval{d(t_1) d(t_4)}  + \expval{d(t_1)} \expval{d(t_4)} \expval{d(t_2) d(t_3)} \\
    & + \expval{d(t_2)} \expval{d(t_3)} \expval{d(t_2) d(t_4)} \left. + \expval{d(t_1)} \expval{d(t_2)} \expval{d(t_3) d(t_4)} \right]  + \mathcal{O}(N^2),
\end{align}
which includes all partitions into two dipole moment expectation values, and a dipole correlation function while preserving the time-ordering of the arguments. 